\documentclass[10pt,authoryear,3p]{elsarticle}
\usepackage{color}
\usepackage{geometry}
\usepackage{graphicx}
\usepackage{amssymb}
\usepackage{amsmath}
\usepackage[round]{natbib}
\usepackage{helvet}

\usepackage[ruled,vlined]{algorithm2e}

\usepackage{color}
\newcommand{\sref}[1]{}

\newcommand{\sifigone}{Supplementary Figure S1 }
\newcommand{\sifigtwo}{Supplementary Figure S2 }
\newcommand{\sifigthree}{Supplementary Figure S3 }

\newcommand{\sifigfive}{Supplementary Figure S4 }
\newcommand{\sifigsix}{Supplementary Figure S5 }
\newcommand{\sifigseven}{Supplementary Figure S6 }
\newcommand{\sifigeight}{Supplementary Figure S7 }
\newcommand{\sifignine}{Supplementary Figure S8 }
\newcommand{\sifigten}{Supplementary Figure S9 }
\newcommand{\sifigeleven}{Supplementary Figure S10 }
\newcommand{\sifigtwelve}{Supplementary Figure S11 }
\newcommand{\sifigthirteen}{Supplementary Figure S13 }
\newcommand{\sifigfourteen}{Supplementary Figure S12 }
\newcommand{\sifigfifteen}{Supplementary Figure S14 }
\newcommand{\sifigsixteen}{Supplementary Figure S15 }

\newcommand{\sitableone}{Supplementary Table S1 }

\newcommand{\pr}[1]{\textcolor{black}{#1}}
\newcommand{\Red}[1]{\textcolor{black}{#1}}
\newcommand{\Green}[1]{\textcolor{black}{#1}}
\newcommand{\Blue}[1]{\textcolor{black}{#1}}
\newcommand{\Sam}[1]{\textcolor{black}{#1}}
\newcommand{\Blueb}[1]{\textcolor{black}{#1}}
\newcommand{\Purple}[1]{\textcolor{black}{#1}}

\newcommand{\beginstar}{%
        \setcounter{table}{0}
        \renewcommand{\thetable}{S\arabic{table}}%
        \setcounter{figure}{0}
        \renewcommand{\thefigure}{S\arabic{figure}}%
}

\usepackage{etoolbox}
\makeatletter
\newif\ifsubsubcaption@ContinuedFloat
\newif\ifsubsubcaption@nonfirst
\patchcmd{\ContinuedFloat}
  {\caption@ContinuedFloat}
  {\subsubcaption@ContinuedFloattrue\caption@ContinuedFloat}
  {}{}

\newcounter{parentsubcaption}

\BeforeBeginEnvironment{subsubcaption}{%
  \ifsubsubcaption@ContinuedFloat
  \else
    \ifsubsubcaption@nonfirst
    \else
      \subsubcaption@nonfirsttrue
      \setcounter{sub\@captype}{0}%
    \fi
  \fi
}
\makeatother

\begin{document}
\begin{frontmatter}

  \title{HyperTraPS: Inferring probabilistic patterns of trait acquisition in evolutionary and disease progression pathways}
  

\author[add1,add3,add5]{Sam F. Greenbury}
\author[add1,add3]{Mauricio Barahona}
\author[add2,add3,add4]{Iain G. Johnston\corref{cor1}}
\ead{iain.johnston@uib.no}
\address[add1]{Department of Mathematics, Imperial College London, UK}
\address[add2]{Department of Mathematics, Faculty of Mathematics and Natural Sciences, University of Bergen, Norway}
\address[add3]{EPSRC Centre for the Mathematics of Precision Healthcare, Imperial College London, UK}
\address[add4]{Alan Turing Institute, London, UK}
\address[add5]{ITMAT Data Science Group, Imperial College London, UK}
\cortext[cor1]{Lead contact}

\journal{Cell Systems}

\begin{abstract}
The explosion of data throughout the biomedical sciences provides unprecedented opportunities to learn about the dynamics of evolution and disease progression, but harnessing these large and diverse datasets remains challenging. Here, we describe a highly generalisable statistical platform to infer the dynamic pathways by which many, potentially interacting, discrete traits are acquired or lost over time in biomedical systems. The platform uses HyperTraPS (hypercubic transition path sampling) to learn progression pathways from cross-sectional, longitudinal, or phylogenetically-linked data with unprecedented efficiency, readily distinguishing multiple competing pathways, and identifying the most parsimonious mechanisms underlying given observations. Its Bayesian structure quantifies uncertainty in pathway structure and allows interpretable predictions of behaviours, such as which symptom a patient will acquire next. We exploit the model's topology to provide visualisation tools for intuitive assessment of multiple, variable pathways. We apply the method to ovarian cancer progression and the evolution of multidrug resistance in tuberculosis, demonstrating its power to reveal previously undetected dynamic pathways.
\end{abstract}

\end{frontmatter}

\section{Introduction}
Many problems in biology, medicine, and throughout the sciences involve the serial stochastic acquisition of discrete features or traits. These traits may be, for example, the symptoms experienced by a patient during progressive disease\Sam{s}, the genetic and physiological features underlying cancer progression, or the acquisition of drug-resistance traits in pathogens. \Green{Understanding the dynamics of these processes has the potential to inform targetted therapies, reveal biological mechanisms, and predict future behaviours, and has been an open challenge throughout the data explosion in biomedical sciences \citep{Colijn2017}.
}

Existing methods to reconstruct the past, and predict the future, of processes involving discrete trait acquisition have emerged from both the cancer science and evolutionary literatures. In the cancer field, disease-related alterations are classified as progressive `hallmarks' \citep{Hanahan2000, Hanahan2011}. Several approaches, reviewed in \citet{Beerenwinkel2015}, utilise computational methods for understanding the way in which cancer progresses via hallmarks at the genetic level \citep{Schwartz2017}. These methods range from stochastic models employing Markov chains for acquisition on graphs such as in \citet{Hjelm2006}, to Bayesian network approaches where trees, forests or directed acyclic graphs (DAGs) are to be inferred from the data \citep{Szabo2002, Beerenwinkel2007, Gerstung2009, Loohuis2014, Ramazzotti2015}. This field often focusses on independent samples exhibiting differing presence of alterations (cross-sectional data) for reconstructing oncogenetic models~\citep{Beerenwinkel2015} to discover progression pathways, or potentially causal relationships between markers, in patients.

Evolutionary and phylogenetic approaches for inferring trait dynamics, by contrast, must account for the relatedness of individuals and the possibility that a given state in a progressive system is inherited from an ancestor. Notable models that have attempted to solve this problem have included Simmap \citep{Bollback2006}, a Markov Chain Monte Carlo (MCMC) approach sampling character mappings on a phylogeny, Ordermutation \citep{Youn2012}, and Reversible Jump MCMC (RJ-MCMC) methodology also applied to a master equation formulation of character dynamics \citep{Pagel2006}. Such approaches have been utilised for understanding the evolution of phenotypic traits in populations \citep{Mahler2010, Watts2015}. In connection with cancer progression, recent modelling approaches aim to reconstructing `phylogenetic' cancer models from sources such as single-cell sequencing data \citep{Beerenwinkel2015, Ross2016, Zafar2017, Ramazzotti2017}. 

Challenges remain in applying these algorithms to dissect the dynamics of systems involving many, potentially coupled, traits. Existing methods may assume a limited number of, or limited interactions between, traits. Computational runtime often scales exponentially with the underlying number of traits, and frequently exhibits challenging scaling with the number of observations. This scaling limits the applicability of some approaches to many forms of biomedical data, particular given modern trends of increasing data volumes and heterogeneity. \pr{Further, several approaches for inferring disease or evolutionary pathways are rather system-specific. In other words, they can process, for example, data on chromosomal aberrations in cancer progression, but are not readily generalised to other (or mixed) data types or diseases. This specificity can be a strength, allowing a more targetted interpretation, but relies on there being specific interest and funding in a particular disease to design a tailored approach for it.}

A recent approach, HyperTraPS (hypercubic transition path sampling) \citep{Johnston2016}, aimed to address these issues, allowing the inference of the dynamics of many coupled traits from \pr{general observational} data following arbitrary (but known) phylogenetic relationships. HyperTraPS represents progressive dynamics as paths on a hypercubic space connecting all possible patterns of trait presence and absence, and uses observations of intermediate states to learn the most likely pathways of progress through this space. In this way, snapshot data can be used to learn the probabilistic structure of dynamic pathways\Green{, which have in turn been used to identify the mechanisms underlying the evolutionary dynamics of $L=65$ mtDNA genes \citep{Johnston2016} and $C_3$ to $C_4$ photosynthesis \citep{Williams2013}.}

\Green{To date, HyperTraPS} has only been used to address these specific evolutionary questions. However, in the current era of large-scale scientific and biomedical data, questions about the structure of dynamic pathways are expanding and becoming increasingly pertinent to evolutionary biology and precision medicine. Hypercubic inference represents a powerful new way of addressing these questions, but a general platform for its application, interpretation, and visualisation remains absent. Such a platform would provide many advantages over the current state of the art: large-scale datasets can be readily analysed, different types of observational data can be used (cross-sectional, longitudinal, and/or phylogenetically coupled observations); Bayesian quantification of uncertainty and a completely unrestricted set of states and transitions can be applied, and competing pathways and their detailed structure can be resolved and characterised, facilitating the identification of progression mechanisms. In principle, any dataset where the relationship of the samples is known or can be inferred is amenable to this detailed analytic approach. 

\Purple{\Green{Here, we address this target,} presenting a novel and expansive set of methodological developments to allow the inference of dynamic pathways from highly general datasets. We embed HyperTraPS in a new and efficient platform for parametric inference and model selection, simultaneously allowing Bayesian inference of dynamic pathways and the identification of model structures that best describe the dynamics and interactions contained within a given set of observations. This model selection simultaneously guards against overfitting and reveal mechanistic insights, namely the extent to which interactions between features dictate the dynamics of the observed system. Models identified in this way have the strongest power to predict out-of-sample observations, which we demonstrate with synthetic and real-world examples, illustrating the predictive power of the approach. To further facilitate interpretation of the inference outcomes, we introduce approaches for intuitively visualising and comparing the high-dimensional pathways inferred from complex datasets, which may include multiple distinct orderings for the acquired traits. \pr{While this overall approach is thus highly general, its Bayesian nature means that domain-specific knowledge constraining a system's behaviour can be readily included for a specific application. This could include, for example, insight into biological mechanisms that forbids feature $A$ appearing before feature $B$, or that suggests the presence of feature $C$ makes feature $D$ twice as likely.}
}

We illustrate the performance of these methods in three different scenarios: with synthetic datasets; with \pr{two datasets on different scales} on the progressive acquisition of genetic alterations in ovarian cancer; and with a recent large-scale dataset on drug-resistant tuberculosis. In these final two cases we demonstrate and discuss several new insights into progression dynamics that the HyperTraPS platform provides. We compare this platform to other approaches \Red{from the disease progression and evolutionary literatures} for trait inference, \Red{highlighting its intersection between these fields and consequent general power and applicability. We conclude by discussing the breadth of applications in the expanding fields of precision medicine, data science, and evolutionary inference, } and provide an open source package for the code.

\begin{figure*}
\includegraphics[width=\textwidth]{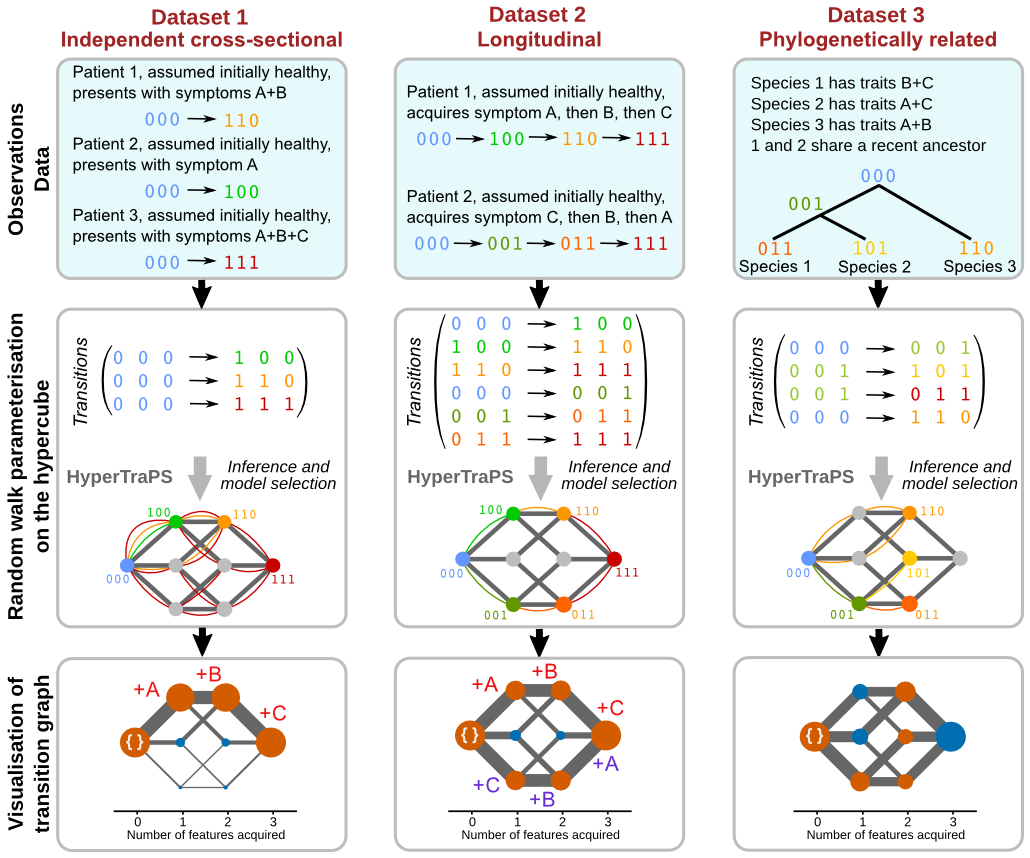}
\caption{\textbf{The HyperTraPS pipeline for learning dynamics underlying cross-sectional and/or longitudinal observations.} \Sam{HyperTraPS allows dynamic inference with three classes of input data. In each case, presence/absence of traits are labelled with a binary marker, and temporal relationships between observations (if present) are invoked to represent observed samples as observed transitions. The likelihood that a given set of edge weights on the underlying hypercubic transition network will give rise to the observed transitions can be calculated efficiently using a path sampling approach (coloured lines). Each illustrative hypercube corresponds to a dataset, with colour coded curved edges and states showing the possible paths that can be taken to reach observed samples. Embedding this likelihood calculation in a Bayesian inference scheme allows posterior weights on inferred transition graphs to be computed, constituting a complete characterisation of the dynamic systems. In the final, visualisation step, the inferred transition graph is embedded and plotted, with edge widths and vertex areas are proportional to the posterior weighting, vertices coloured according to whether they reflect observed (orange) or hidden (blue) states, and paths labelled by the progressive acquisition of features.}
}
  \label{fig:methods}
\end{figure*}

\section{Results}

\subsection{Inferring dynamic pathways involving coupled traits on general state spaces}\label{section:csd}
HyperTraPS represents every possible state of a system with $L$ features or traits (we use these terms synonymously here) as a binary string of length $L$, where 0 and 1 at the $i$th position correspond respectively to absence or presence of the $i$th trait. Traits are acquired stochastically and irreversibly, according to transition probabilities linking states on a \Red{hypercubic transition graph} (Fig. \ref{fig:methods}). \Red{We consider instances of an evolving or progressing system as an ensemble of random walkers on this graph. As in a hidden Markov model \Purple{\cite{Murphy2012}}}, observations are assumed to arise through signals randomly emitted by \Red{these walkers}; a signal corresponds to the current set of acquired traits of the random walker. The task at the core of HyperTraPS is to compute the likelihood of observing a set of emissions that match the transitions in a dataset, given a parameterisation $W$ describing the transition probabilities on the edges of the hypercube.

In Methos Appendix, Fig. \ref{fig:methods}, and \sifigone, we outline the HyperTraPS algorithm to estimate this likelihood given a set of observations. As Fig. \ref{fig:methods} illustrates, these observations can be independent and cross-sectional (for example, single snapshots of symptom presence/absence in independent patients), longitudinal (for example, time series of symptom presence/absence in the same patients over time), and/or phylogenetically related (for example, evolving traits which may be inherited from ancestor to descendent). \pr{Cross-sectional and longitudinal data structures involve many independent evolutionary processes running in parallel; phylogenetic data structures involve an initially single process that may branch, with different branches subsequently evolving independently.} \Green{In contrast to previous approaches \citep{Johnston2016, Williams2013}, we embed the core likelihood calculation in an auxilary pseudo-marginal MCMC (APM MCMC) framework \citep{Murray2015} to allow more efficient Bayesian inference of the hypercubic transition network supporting the observed dynamics. The APM MCMC embedding overcomes potential issues arising from uncertainty in the likelihood estimates \Sam{for} long pathway calculations (Methos Appendix), better guaranteeing that the MCMC process will mix well and converge to a consistent posterior in the case of large, sparse inference challenges. \pr{For example, in the ovarian cancer inference presented below, the APM embedding reduced the characteristic MCMC mixing time by a factor of 5.} APM MCMC makes it possible to address systems involving dozens of sparsely sampled traits, as we demonstrate below.} 

\newcommand\BS{\begin{subfigure}[t]{0.32\textwidth}}
\begin{figure*}[!tbh]
  \centering
   \includegraphics[width=.9\textwidth]{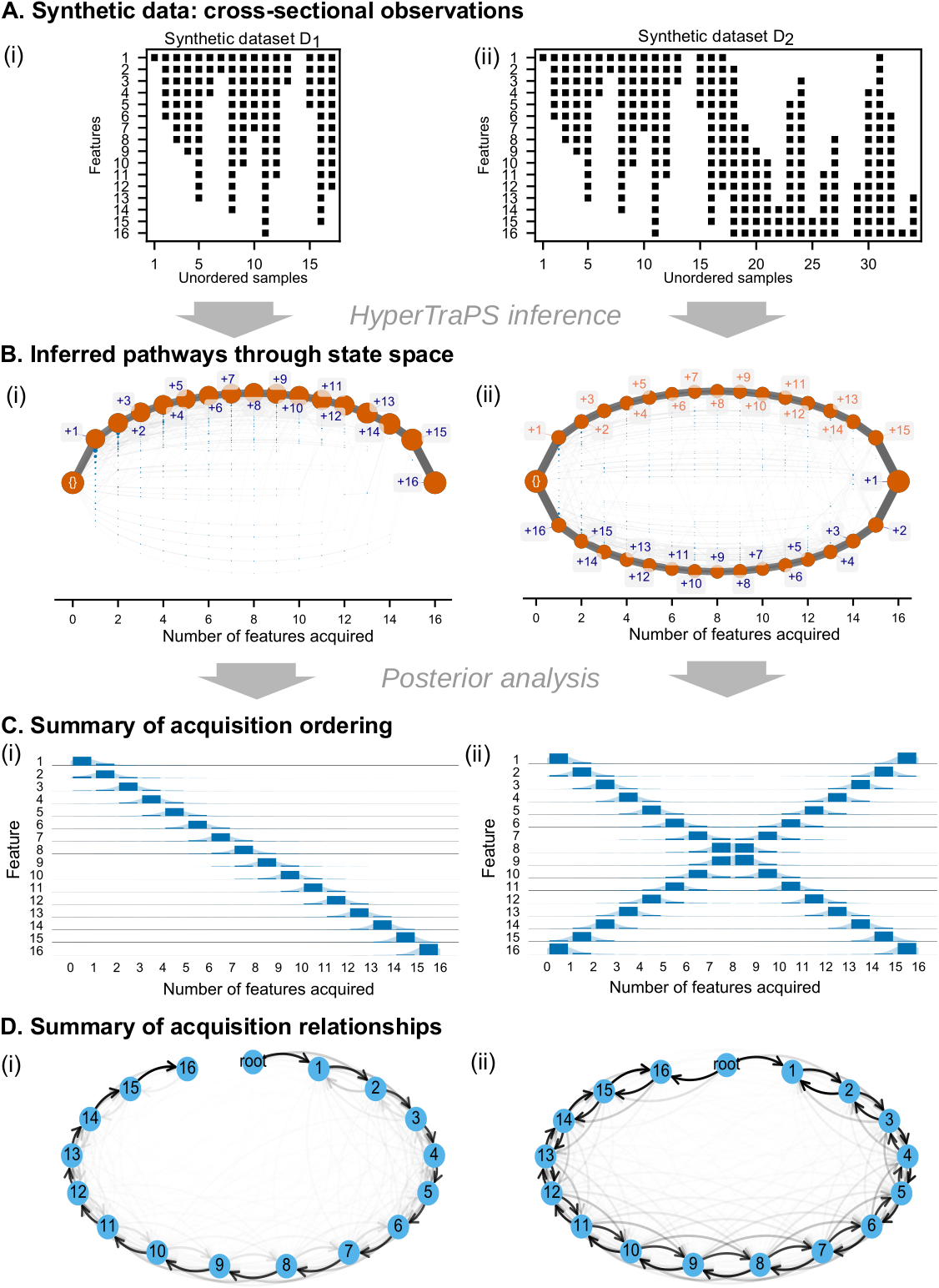}
 	\caption{\textbf{HyperTraPS inference for two synthetic datasets.}
	\textbf{(A)} Synthetic datasets used in the inference process. (i) Dataset $D_1$ supports only a single pathway; (ii) dataset $D_2$ supports two competing pathways with features acquired in opposing orders.
	\textbf{(B)} Inferred dynamics on the hypercubic transition graph. Edge widths and node areas are proportional to the number of times edges/nodes are encountered. States are plotted from left to right in order of the number of features acquired (embedding and labelling procedure described in STAR methods). The single pathway clearly dominates in (i), while the two competing pathways are clearly observable in (ii).
	\textbf{(C)} Inferred dynamics represented as the posterior probability that a feature (horizontal axis) is acquired at a given step (vertical axis). Bimodality in ordering posteriors (ii) reflect the presence of distinct progressions that exist in the underlying dynamics.
	\textbf{(D)} Inferred dynamics represented as a \pr{directed} graph \pr{(edges run from left to right in these embeddings)} summarising trait acquisition relationships to the previous acquisition.
	Paths on these graphs reflect possible acquisition ordering inferred by HyperTraPS: respectively a single pathway (i) and two pathways (clockwise and anti-clockwise) in opposite directions (ii).}
	\label{fig:cs12}
\end{figure*}

The next important consideration in this inference process is how this transition network is parameterised. Individually parameterising each of $L 2^{L-1}$ hypercubic edges represents a substantial inference challenge for (likely) very little model fit reward. Instead, we propose a hierarchy of parameter representations (\sifigtwo \pr{; Methos Appendix}). For the \emph{zero order} model every feature has equal probability of acquisition. All edges on the transition network thus have the same weight, requiring no parameters. In the \emph{first order} model, every feature has an independent acquisition probability regardless of current state. Transition edge weights between two states are thus exclusively determined by the trait that distinguished the two states (requiring $k=L$ parameters). In the {\em second order} model, every feature's acquisition probability depends independently on the presence of each other feature. Transition edge weights between two states thus depend on the distinguishing trait and the presence/absence of each other trait (requiring $L^2$ parameters; as in \citep{Johnston2016}). Higher order models, including the full $L 2^{L-1}$ set \Red{naturally follow}, introducing more complex interactions between the co-occurrence of features (as in \cite{Williams2013}). The appropriate choice of parameterisation is dictated by the generative processes underlying the observed data; if trait acquisitions are independent, the parsimonious first-order model is more appropriate; if traits interact pairwise, the second-order model will be required to capture the dynamics. A given dataset may be best described by an intermediate representation between two of these cases.

To identify the optimal parameter representation for a given dataset, we introduce methods for regularising the inferred model parameterisations (see Methos Appendix), allowing the
\Purple{appropriate choice of model structure to describe the observed data and a means of generating maximum likelihood parameterisations without overfitting.}
As we demonstrate below, the regularisation process allows us to distinguish simple cases, where all dynamics can be described by traits behaving independently, from more complex cases where the acquisition of one or more traits influences \Sam{the probability of acquisition of other traits}. This combination of an efficient and general inference platform, a process for model selection, and a new toolbox for visualising and interpreting inferred posteriors, allows us for the first time to apply HyperTraPS to a dramatically expanded range of biomedical questions.

\subsection{Inference of pathways from synthetic data}
To illustrate the ability of HyperTraPS to characterise dynamics from independent cross-sectional samples, we constructed two cross-sectional datasets with different underlying progressions. The first ($D_1$; Fig. \ref{fig:cs12}A(i)) involves samples taken uniformly from each state along a single trajectory, where features are accumulated from {\em left to right}. For example, for $L=3$, the sequence of acquisition would be $000 \rightarrow 100 \rightarrow 110 \rightarrow 111$. The second ($D_2$; Fig. \ref{fig:cs12}A(ii)) involves samples taken uniformly from states along two distinct progression pathways with exactly opposing temporal ordering of acquisition: one where features are acquired from {\em left to right} and the other where features are acquired from {\em right to left}. For example, for $L=3$, this would correspond to the two trajectories $000 \rightarrow 100 \rightarrow 110 \rightarrow 111$ and $000 \rightarrow 001 \rightarrow 011 \rightarrow 111$.

\Blue{We chose these structures to illustrate HyperTraPS' ability to infer both single and multiple competing pathways}. \Red{For the single pathway, traits can be independent -- a suitable ordering of the `basal rates' is sufficient to generate the observations.} \Red{By contrast, competing pathways require traits to interact -- acquisition of traits on one pathway must repress acquisition of traits on the other pathway}. 

\Blue{In Fig. \ref{fig:cs12}, we show the structure of the data and the outcomes of the inference process. To visualise the learned dynamic behaviour, we use a customised algorithm (described in further detail in STAR methods) to project the inferred hypercubic transition network into two dimensions, arranging states with increasing numbers of features from left to right (Fig. \ref{fig:cs12}B). A single dominant progression is clear for Fig. \ref{fig:cs12}B(i), while the two progressions are clearly shown in Fig. \ref{fig:cs12}B(ii). Fig. \ref{fig:cs12}C shows an alternative representation: the posterior probabilities with which each trait is acquired in each possible ordering. Again, the dynamics corresponding to the simple single pathway and the more complex competing two-pathway model are clearly visible.}

\Blue{In this extreme example, the inferred ordering distributions for all but the central traits in the multiple-pathway case (ii) exhibit {\em bimodality}. Generally in such histograms from HyperTraPS posteriors, bimodality (and multimodality more generally) reflects structurally distinct progression pathways (for example, where a feature can be acquired early or late, but not at intermediate stages), while unimodal distributions reflect sets of pathways with a consistent structural trend. The width of such modes reflects the amount of variability in the order for which a feature is acquired in the progression associated with the mode. Multimodal distributions in these plots provide a suggestive signature of distinct dynamic pathways of the system. \Red{In Methos Appendix and \sitableone, we compare this inference of competing pathways to existing alternative approaches and show that HyperTraPS has a unique ability to resolve and characterise multiple progressive pathways.}}

In Fig. \ref{fig:cs12}D, we represent dynamics from the inference process as {\em probabilistic feature graphs} \pr{(PFGs)}, allowing more direct comparison with existing approaches. These PFGs summarise the probability the feature $Y$ is acquired next, given that feature $X$ was last to be acquired (see Methos Appendix). Once more, in Fig. \ref{fig:cs12}D, the single monotonic path in (i) and two paths for (ii) are clearly visible.

\pr{In Fig. \ref{mtnewfig} and \sifigthree - \sifigsix, we demonstrate the performance of the inference process under availabilities and structures of source data, and in the presence of prior knowledge about pathways. \sifigthree shows that characteristic pathways can readily be identified under each of the three different types of data from Fig. \ref{fig:methods}. The resulting posterior distributions are sharper for cross-sectional data than for longitudinal and phylogenetic data, reflecting the fact that the independent samples from cross-sectional data provide more evidence for corresponding pathways than the coupled data in the other cases. Fig. \ref{mtnewfig}A shows the ability of HyperTraPS to identify pathways given limited data ($N = 10$ observation are sufficient to broadly characterise a single pathway for an $L = 16$ system; $N = 50$ gives near-perfect reconstruction). Even for competing pathways, $N \geq 20$ serves to provide information on pathway structure in this case. Fig. \ref{mtnewfig}B and \sifigfive demonstrates that HyperTraPS can readily discern several completely independent pathways (8 pathways can be readily identified for the $L=16$ system; 16 completely independent pathways pose more of a challenge). Finally, Fig. \ref{mtnewfig}C and \sifigsix highlight the Bayesian nature of HyperTraPS by demonstrating how the inclusion of prior knowledge about pathway structure can help resolve degeneracy in the identified solutions, for repeated and/or incomplete observations. To summarise, HyperTraPS can readily identify pathway structure including multiple, competing, independent pathways, using limited volumes of data, and can readily harness prior knowledge.}

\begin{figure*}[!tbh]
  \centering
   \includegraphics[width=1.\textwidth]{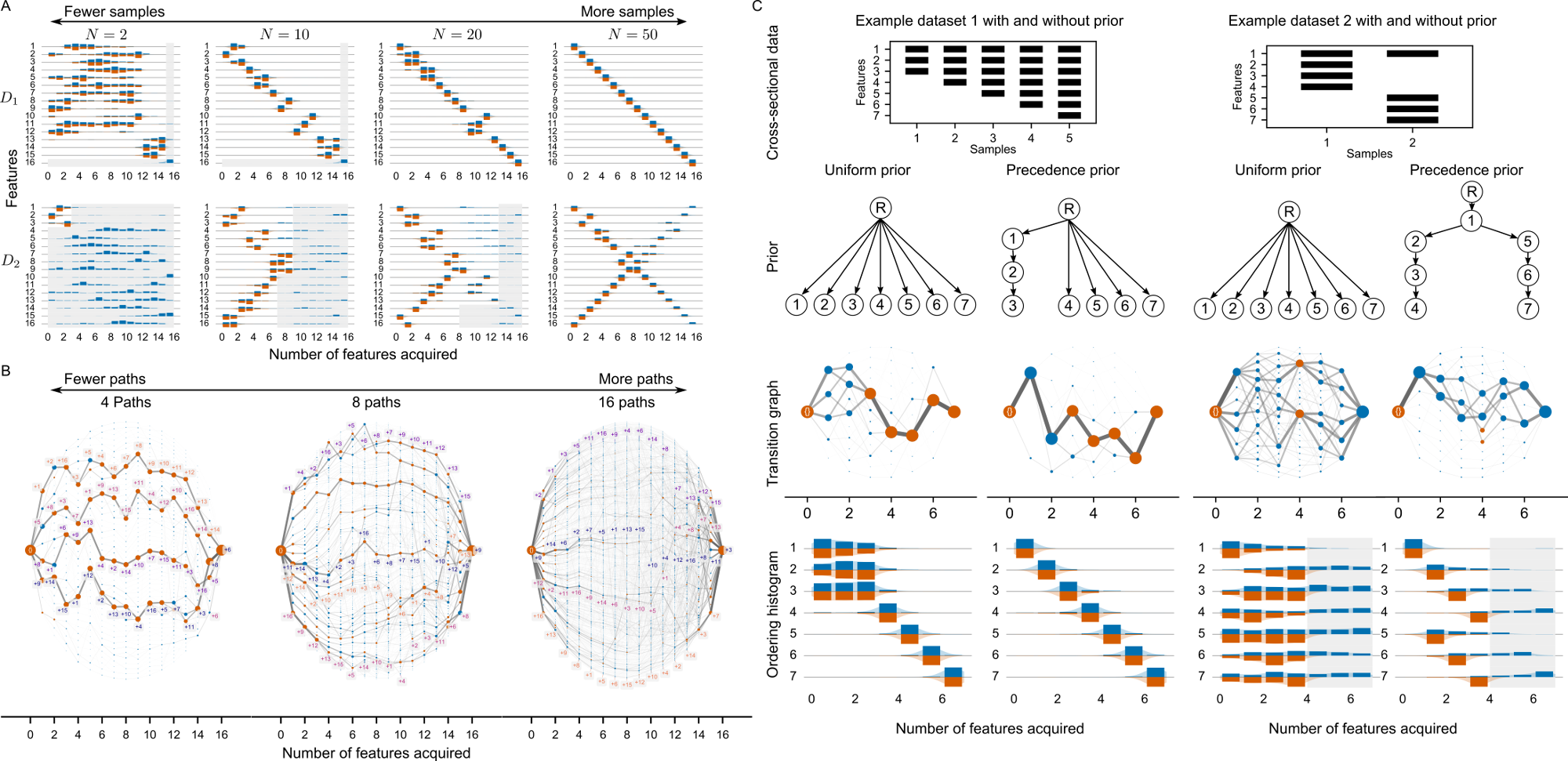}
 	\caption{\textbf{HyperTraPS inference of pathway structures under different conditions.} \textbf{(A)} Inference of $D_1$ (single pathway) and $D_2$ (two competing pathway) structures with $L = 16$ and increasing number of sample states $N$. Both pathway structures are readily identified for $N \geq 20$. Lower $N$ challenges reconstruction of pathways, although the outline of the single pathway is still visible for $N = 10$. At lower $N$, the posteriors tend towards the uniform priors. \textbf{(B)} Inference of $p$ competing pathways with $L = 16$ and $N = 16p$ samples. 8 completely distinct pathways are readily identified with clearly distinguished posterior density; 16 independent pathways pose more of a challenge but are still identified. \textbf{(C)} Including prior information in HyperTraPS inference. (left) Single pathway dataset without any observations with fewer than three acquisitions. Inference without prior information of these first three features leads to a uniform inference of acquisition order; including a prior tree (see text) recovers the true ordering. (right) Two competing trees of acquisition order for prior information. Without prior information there is large heterogeneity in the order and precedence of feature acquisition following inference. Including prior information canalises the inferred pathways and recovers the original structure.}
	\label{mtnewfig}
\end{figure*}

\pr{This final point is particularly pertinent when applying HyperTraPS to specific scientific questions. When uninformative priors are used, HyperTraPS is a highly general approach, where mechanistic inference is guided by the data alone. For domain-specific cases -- for example, particular diseases, or particular metabolic pathways -- subject-specific knowledge may constrain the allowed pathways (for example, mechanistic insight may forbid or favour transitions between particular states). In these cases, the inclusion of this knowledge via prior distributions as in Fig. \ref{mtnewfig}C and \sifigsix can readily and generally be used to constrain the posterior dynamics supported by HyperTraPS.} 

In Methos Appendix and \sifigseven - \sifigtwelve, we further expand upon these test cases (\sifigseven - \sifigeight) and the interpretation of pathway dynamics (\sifignine), and demonstrate that HyperTraPS successfully learns pathways in the case of partial (\sifigten), noisy (\sifigeleven), and non-uniform (\sifigtwelve) sampling.

\subsection{Model regularisation and validation}\label{section:csdxxx}

We next demonstrate how regularisation can be used to determine the optimal model structure required to describe and predict features of the two synthetic datasets. $D_1$ is produced by a model with no trait interactions, and hence requires only $L$ independent parameters to reproduce its dynamics. $D_2$ requires interactions between traits: progress along one pathway must suppress progress along the other. More parameters are thus required to encode these interactions to adequately match the data. We therefore asked if, given a range of starting model representations, the regularisation process could identify the appropriate number of parameters for each case.

\Sam{Fig. \ref{fig:mt_regularisation}A demonstrates this regularisation process. For $D_1$, the first-order model remains intact with its original $L$ parameters, and the second-order model is reduced from $L^2$ to $\sim L$ parameters, reflecting the fact that $L$ (and only $L$) parameters are required to capture the single-pathway dynamics. This regularised second-order model performs equally well to the first-order model.}

For $D_2$, with two competing pathways, the first-order model fails to capture the observed behaviour even with its full set of $L$ parameters. The regularisation process reduces the first-order model to $\sim 0$ parameters: as no instance of model 1 can adequately describe the observations, the parameter set is minimised for parsimony. By contrast, the second-order model is reduced to $\sim 2L$ parameters, which provides an optimal description of the data. The requirement for higher-order terms here is a consequence of the trait-interaction terms in the second-order model allowing the required cross-repression of pathways, making it a better explanatory model in this case.

\Blue{To validate these findings and explore the predictive power of our inference platform, we split the data into two halves to form a training and test dataset.
We obtained posteriors from the training set for each model, and computed the likelihood associated with the test set for these inferred posteriors. \Sam{Fig. \ref{fig:mt_regularisation}B} shows the AIC scores for the full model, and the log-likelihoods for training and validation datasets. For the single-pathway dataset $D_1$, the first order model and second order model provide similar explanatory power in the full model, and predictive power in the validation experiment, both improved over the zero order model (null model). For the two-pathway dataset $D_2$, the second order model enhances predictive power compared to both the first order and null models, and regularisation improves the parsimony of this model with no cost to model fit ($p < 0.001$ for the a likelihood ratio test against the null model).} 

\begin{figure}[!tbh]
  \centering
    \includegraphics[width=.8\textwidth]{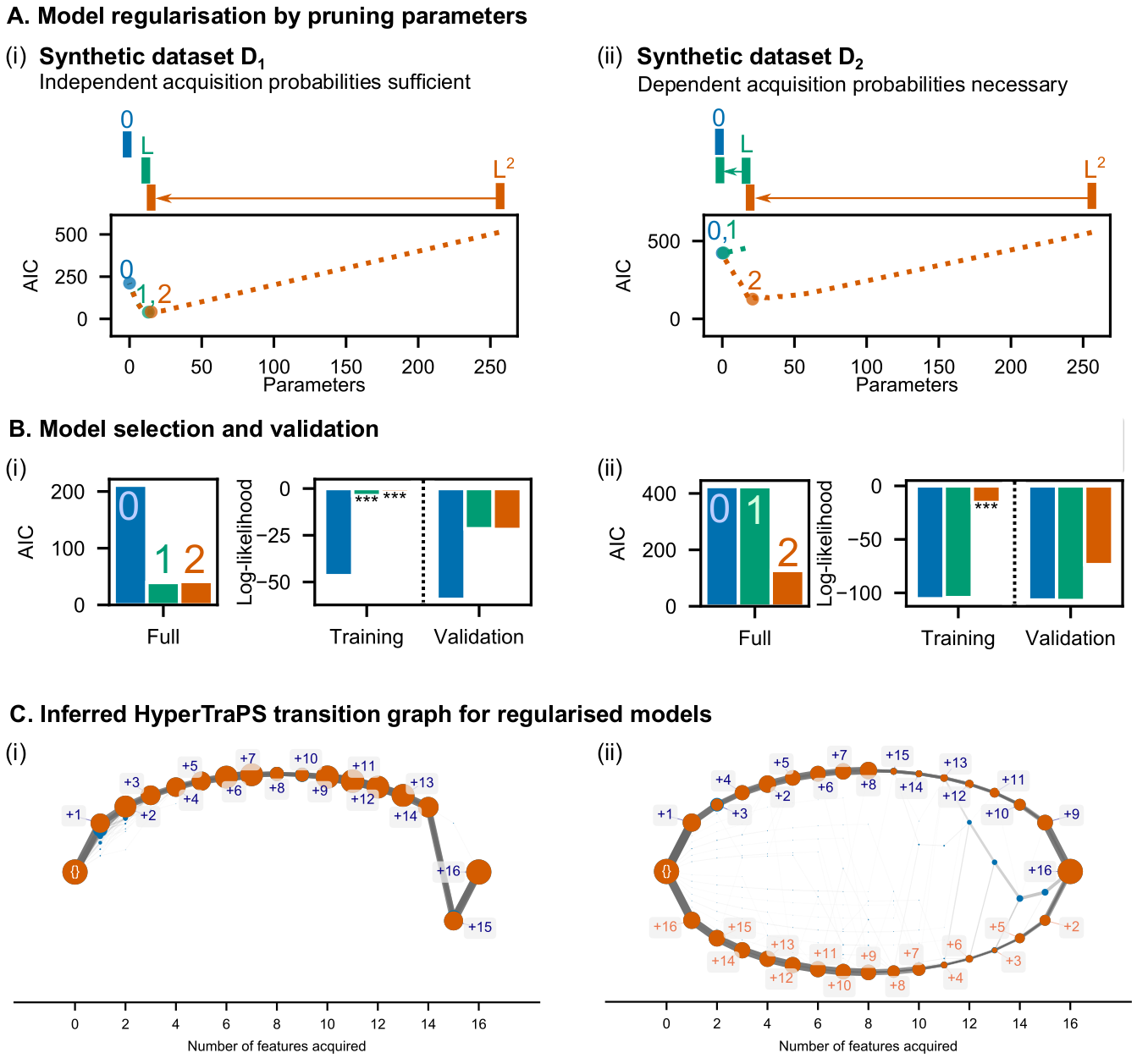}
    \caption{\Blueb{\textbf{Regularisation and model comparison for HyperTraPS inference.} We compare models `0' (zeroth order, 0 parameters, all traits are acquired with the same independent probability); `1' (first order, $L$ parameters, acquisition probabilities are independent but may differ); and `2' (second order, $L^2$ parameters, pairwise interactions between trait acquisition probabilities) for the datasets $D_1$ and $D_2$ in Fig. \ref{fig:cs12}. \textbf{(A)} Model regularisation. Parameters are greedily pruned from each inferred model to identify a reduced parameter set that minimises AIC. The turning points illustrating an optimally sparse parameterisation are marked for each model. \textbf{(B)} Model selection and validation. (left) AIC scores for the regularised version of each model; (right) likelihoods for the training and validation datasets (see text). For the full and training dataset, stars give the $p$-value from a likelihood ratio in comparison to the zero order model (the null model) with significance levels of ${*} {*} {*} < 0.001$, $\ast \ast < 0.01$ and $\ast < 0.05$. \textbf{(C)} Inferred dynamics on the hypercubic transition graph for the regularised first order model for $D_1$ and for the regularised second order model for $D_2$. Each corresponding pathway is still well captured despite substantial parameter reduction.}}
    \label{fig:mt_regularisation}
\end{figure}

\subsection{Comparison with existing inference approaches}
\begin{figure*}[!tbh]
  \centering
    \includegraphics[width=\textwidth]{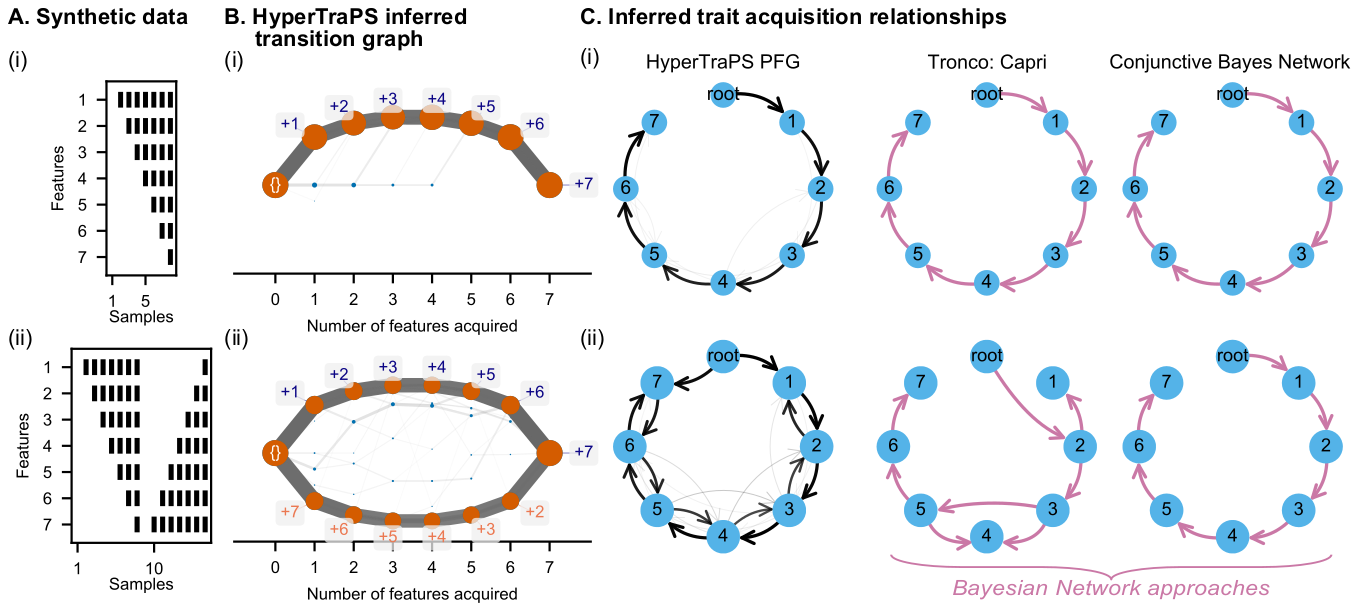}
    \caption{\textbf{Comparison between HyperTraPS and alternative inference approaches.} \textbf{(A)} Synthetic datasets (i) and (ii) following the forms from Fig. \ref{fig:cs12}. \textbf{(B)} Full inferred transition graphs from HyperTraPS. \textbf{(C)} Inferred transition graphs from HyperTraPS represented as probabilistic feature graphs, compared to alternative inference approaches. All approaches agree for the simple, single pathway (i). For the competing pathways (ii), the full inferred transition graph from HyperTraPS, and its structure summarised into trait contingencies, capture the two alternative pathway structures, while alternative approaches (highlighted) are more challenged, either presenting a combination of steps from both pathways or exclusively reporting one.}
    \label{fig:synthetic_feature_graph}
\end{figure*}

We next sought to compare the outputs of the HyperTraPS inference process to existing approaches to infer dynamic pathways from data (Fig. \ref{fig:synthetic_feature_graph}). We highlight here that HyperTraPS is, to our knowledge, the only inference approach that attempts to learn the transition rates (with uncertainties) between every possible state of a system. Other approaches typically focus on a reduced subset of states. The full, high-dimensional posteriors inferred by HyperTraPS therefore cannot be readily compared with the outputs of other approaches. However, summaries of these posteriors, losing some information, can more naturally be compared with lower-dimensional alternatives.

To this end, we compared reduced summaries of the dynamics learned by HyperTraPS with the Bayesian networks derived from the Capri algorithm \cite{Ramazzotti2015} and Conjunctive Bayes Network approaches \citep{Montazeri2016} (using MC-CBN, the most recent CBN package for large or small scale inference), two commonly used Bayesian network methods in the literature, using synthetic datasets (Fig. \ref{fig:synthetic_feature_graph}). \Blueb{These approaches produce directed acyclic graphs (DAGs) on the set of features, where an edge between $X$ and $Y$ denotes an inferred causal relationship between $X$ and $Y$. Such representations allow for possible causal relationships between features to be found, but {\em a priori} impose that such relationships exist and are monotonic. For example, if trait $X$ influences the presence of trait $Y$, trait $Y$ may not influence trait $X$. Overall ordering of feature acquisition may not be unique (a joint probability distribution of events may have underlying degeneracy in the order of those events), but the monotonic relationship between features does impose partial ordering. In HyperTraPS, no monotonic precedence is imposed between features: $X$ may influence $Y$ and $Y$ may influence $X$. This relaxation allows, for example, cross-repression of traits, as we shall see for dataset $D_2$. For comparison with other approaches, we condense the full output of inference (DAGs in state space, i.e. on the hypercube) into graphs in feature space.}

\Blueb{As seen in Fig. \ref{fig:synthetic_feature_graph}, for the single-pathway case of dataset $D_1$, all graphs have the same structure and therefore are in agreement over the single pathway that most likely explains the data. For the competing pathway case of dataset $D_2$, the outputs are different in each case. The HyperTraPS feature graph captures the dual pathways, with directed edges between each pair of non-root nodes. Capri is unable to resolve a meaningful relationship between features, because the competing pathways frustrate the assignment of temporal priority between the features. The outputted graph is therefore unable to recover a significant relationship between features representing precedence relationships.} The Conjunctive Bayes Network is able to resolve one of the directed paths but not the other.

These comparisons have been performed with cross-sectional synthetic observations. As discussed above (Fig. \ref{fig:methods}A), HyperTraPS can also infer dynamic pathways given longitudinal and phylogenetically coupled data. Ref. \cite{Johnston2016} demonstrated that HyperTraPS has several advantages over existing approaches for trait inference on phylogenies. In Methos Appendix and Discussion, we pursue these comparisons further and show that HyperTraPS presents several scaling and performance advantages over alternative methods, again reflecting its ability to resolve independent pathways involving many coupled traits.

Taken together, these results provide support for our platform's ability to learn single progression pathways efficiently and also dissect competing progression pathways more directly than alternative approaches. We reiterate that, in addition to these coarse-grained readouts, HyperTraPS learns explicit probabilities for transitions between every state of a system, allowing a still finer resolution of dynamics.

\subsection{Application to cross-sectional ovarian cancer data}

\begin{figure}[!t]
  \centering
  \includegraphics[width=\linewidth]{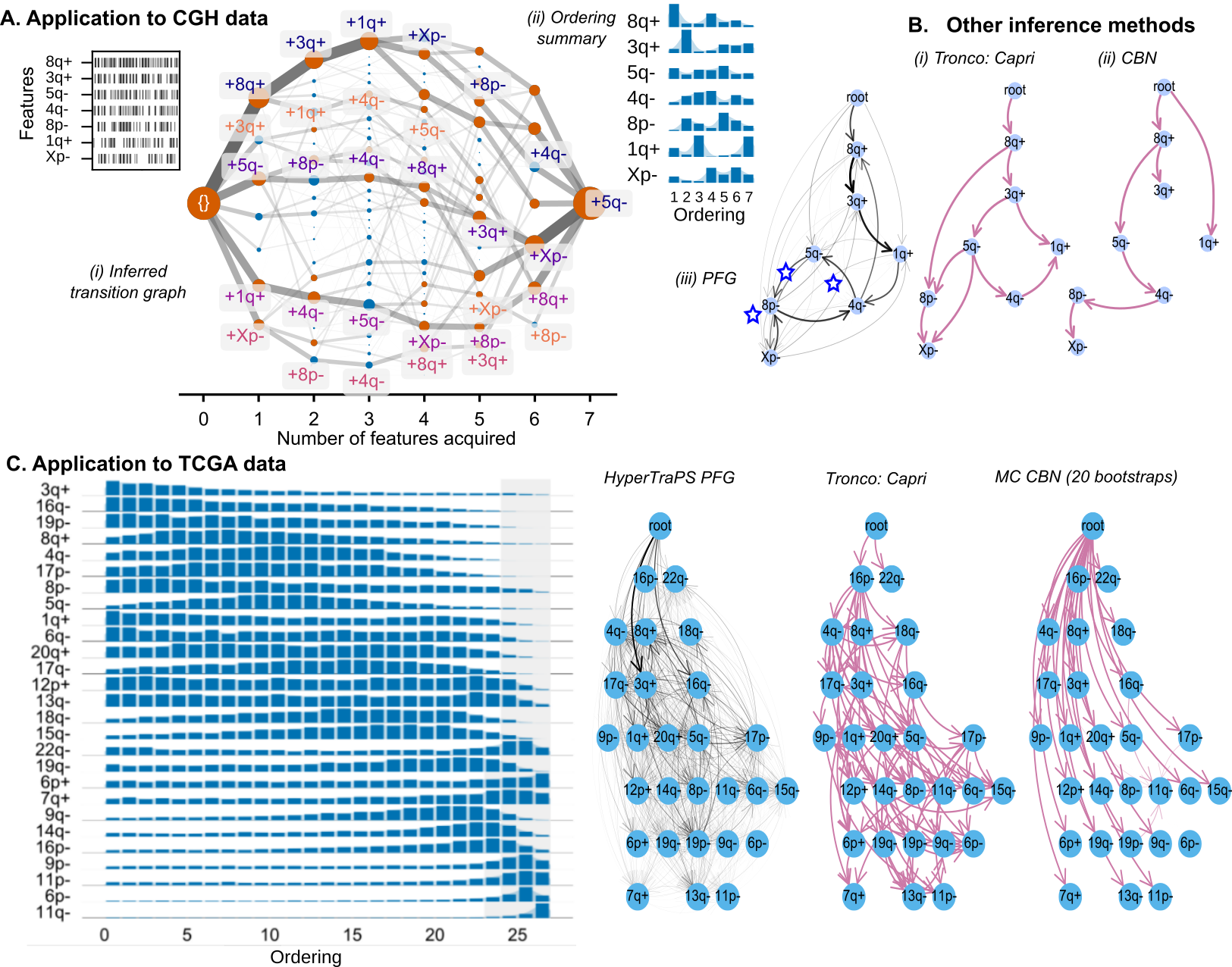}
  
        \caption{\Blueb{\textbf{HyperTraPS inference for ovarian cancer progression reveals canalised progression pathways and new transition information.}
	\Blueb{
        \textbf{(A)} \pr{HyperTraPS inference applied to a dataset (inset) of cross-sectional observations of chromosomal aberrations in an ovarian cancer dataset. The inference process produces} transition graph (i), \pr{summary ordering posterior (ii), and corresponding probabilistic feature graph (iii)}, reflecting the inferred dynamics of cancer progression. Progression pathways are substantially canalised, with the first acquired aberration feature largely determining the subsequent dynamics of the disease. Starred edges in \pr{(iii)} correspond to edges present in the probabilistic feature graph, related to the $4q- \rightarrow 5q-$ system discussed in the text, that are absent in other approaches.
        \textbf{(B)} Trait relationships inferred with alternative computational approaches. HyperTraPS largely agrees with the core structure of alternative approaches (especially with Capri where there is less strict constraints on precedence) but reveals several additional features (illustrated with stars in (B)). For example, the $4q- \rightarrow 5q-$ pathway from is omitted and directly opposed in alternative approaches where only monotonic relationships between {\em 5q-} and other features are permitted. Further, the canalised structure present in B is not naturally captured by the inferred outputs of the alternative approaches.
        \pr{\textbf{(C)} Inferred orderings of chromosomal changes in ovarian cancer progression using observations from the cancer genome atlas (TCGA) dataset, and corresponding inferred transition graphs from the TCGA inference compared to alternative approaches as in (B).}
        }}}
  \label{fig:ovarian-cgh}
\end{figure}

\Blueb{To demonstrate HyperTraPS' ability to elucidate dynamic pathways of biomedical importance, we next asked whether our approach could be used to infer pathways of cancer progression. The field of cancer progression models is diverse, with many methods designed for performing inference with different types of data \citep{Beerenwinkel2015, Schwartz2017}. As \citet{Schwartz2017} discuss, data relating to alterations in cancer broadly belong to three categories: bulk tumour samples from different patients, bulk tumour samples from different tumours within a single patient, or single cell data typically from a single tumour. Computational methods can broadly be categorised into those inferring the phylogenetic relationship of samples (their history and genealogy), and those inferring direct relationships between the features suggestive of precedence or progressions relating to feature acquisitions. We discuss the methods within the cancer progression model literature further in Methos Appendix.}

As illustrated in Fig. \ref{fig:methods}, HyperTraPS can both handle independent and arbitrarily dependent samples, and so can be used with any of the above types of dataset. \Blueb{We here focus on the case of independent bulk samples from different patients where there is no phylogenetic relationship between samples, as it is assumed that features are acquired during a patient's lifetime.} Existing approaches for this problem \citep{Beerenwinkel2015} focus on the reconstruction of different types of Bayesian network relating the acquisition of genetic alterations relating to the progression of cancer. As cancer is directly related to the acquisition of driver mutations that provide fitness advantage for the cells in which they are acquired, recent work such as \citet{Diaz-Uriarte2018} has argued for the need to consider cancer progression from a different perspective in which features may have multiple orderings due to the high-dimensional structure of fitness landscapes and the potential presence of epistatic effects. The HyperTraPS platform directly allows this inference of multiple paths. 

\Blueb{We first applied HyperTraPS to the well-studied dataset for chromosomal alterations in ovarian cancer, recovered through Comparative Genomic Hybridization (CGH) \citep{Knutsen2005}. This dataset is included in the Oncotrees package \citep{Szabo2002} and utilised in comparisons with the Caprese algorithm \citep{Loohuis2014}. The data consist of a sample of $N=87$ patients for $L=7$ chromosomal alterations associated with ovarian cancer, with the assumption that none of the alterations were present in the individual at birth.}

Fig. \ref{fig:ovarian-cgh}A and \sifigfourteen provide a visual representation of the dataset, showing the presence/absence of each genetic alteration in each patient. Fig. \ref{fig:ovarian-cgh}A(i) shows the recorded transitions following parameter inference on the hypercube. A set of several constrained, well-defined paths are visible, with flexible ordering in the acquisition of initial features being apparent. Interestingly, the feature that is acquired first has substantial influence over the subsequent pathway structure, visible as the tightly constrained individual pathways in Fig. \ref{fig:ovarian-cgh}A(i) with rather few transitions between pathways, \pr{and as bimodal structure in the posterior summary plot in Fig. \ref{fig:ovarian-cgh}A(ii)}. This canalisation suggests substantial memory effects in the later stages of cancer progression.


\Blueb{To further examine the multiple non-monotonic pathways that the data may contain, we make use of the probabilistic feature graphs described above and in Methos Appendix. Fig. \ref{fig:ovarian-cgh}A(iii) shows the probabilistic feature graph between each pair of features and Fig.~\ref{fig:ovarian-cgh}B shows Bayes network representations of feature relationships from alternative approaches. Here, as above, each edge is directed and has a weight in proportion to the probability of acquiring feature $Y$ having just acquired feature $X$. Elements of the core structure are shared between the HyperTraPS, Capri, and CBN approaches.}

\Blueb{To demonstrate another example of where HyperTraPS' increased detail allows new insight into multiple pathways, we focus on several transitions that have strong edges in the HyperTraPS PFG that are missing from the other approaches. In both alternative Bayesian network approaches, an edge is present from {\em 5q-} to {\em 4q-} but never the other way around. The precedence in these models is due to the fact that {\em 5q-} is more frequent than {\em 4q-} and the need to ensure a monotonicity between features to construct the desired Bayesian network output. As HyperTraPS places no such restriction, it is capable of finding additional pathways in which {\em 4q-} is acquired prior to {\em 5q-}. As seen in Fig. \ref{fig:ovarian-cgh}B(i), this ordering may be achieved in several ways through the acquisition of {\em 8p-}, {\em 3q+} or {\em 1q+} and, given the acquisition of those features, is in fact more likely to be acquired prior to {\em 5q-}. The acquisition of \emph{4q-} prior to \emph{5q-} is indeed observed in 10 of 87 (11.5\%) samples in the data.}

\pr{Having gained substantial insight from this comparatively simple dataset, we next asked whether HyperTraPS could be used with the larger volumes of data that emerge from more modern genome-scale studies. To this end, we obtained raw data from the cancer genome atlas (TCGA) project \citep{Bell2011}. We converted these raw data into feature `barcodes' over a variety of scales, yielding a set of cross-sectional datasets (see Methos Appendix). First, we constructed a dataset describing the chromosomal regions in which each of the $N=489$ patients had amplifications/deletions above the significance threshold defined in the study. This gave a dataset describing each patient's presence or absence of aberration in $L=55$ regions. Secondly, we considered the subset of the $L=27$ chromosomal regions marked as of particular interest in Fig. 1c of \citet{Bell2011}.
}

\pr{Our APM MCMC embedding of HyperTraPS allowed the algorithm to readily produce posterior distributions in each case. In the first, larger, case, posteriors show a clear ordering in the acquisition propensity for different chromosomal features (\sifigthirteen). However, the large dynamic space associated with these $L=55$ features makes more detailed interpretation of these posteriors rather laborious. This reflects a challenge in the application of HyperTraPS: while posteriors can readily be obtained for large numbers of features \citep{Johnston2016}, the interpretation of these posteriors can be challenged by the output volume.}

\pr{Consistent with this, the results from the subset of regions are more interpretable (Fig. \ref{fig:ovarian-cgh}C). Here, clearly converged posterior distributions are visible, with some bimodality (for example, in features \emph{1q+, 13q-} and \emph{22q-}) suggesting the presence of competing pathways. In particular, bimodality in the \emph{1q+} posterior reflects the multiple associated pathways in the previous CGH dataset (Fig. \ref{fig:ovarian-cgh}A). The orderings of other features from the CGH dataset are consistently reflected in HyperTraPS' treatment of the TCGA data, with the additional volume of data in the TCGA case helping to further detail posterior structure. Interpreted as a PFG (Fig. \ref{fig:ovarian-cgh}C), these posteriors highlight both the heterogeneity of, and strong structures within, the associated progression pathways. Strong early edges, for example, surround the \emph{3q+} feature, linking $\emptyset \rightarrow$ \emph{3q+} and \emph{3q+} $\rightarrow$ \emph{8q+}, and the \emph{16q-} feature.}

\pr{Other approaches do not capture several of these transitions. For example, 70 samples in the dataset possess the $3q+$ feature but not $8q+$ (compared to 71 which possess $8q+$ but not $3q+$), while the Capri Bayesian network is only able to identify a single causal relationship from $8q+\rightarrow 3q+$ and the CBN approach does not identify any edge between the pair (due to this large proportion of conflicting samples).}


\pr{These biomedical examples serve} to illustrate the power of the HyperTraPS to infer multiple competing pathways providing interpretable representations of such paths, and further the shortcomings of alternative approaches that restrict the output of learnt networks to be of the Bayesian network variety.

\subsection{Application to the evolution of multi-drug resistant tuberculosis}\label{section:tb}
\begin{figure*}[!t]
  \centering
  \includegraphics[width=\textwidth]{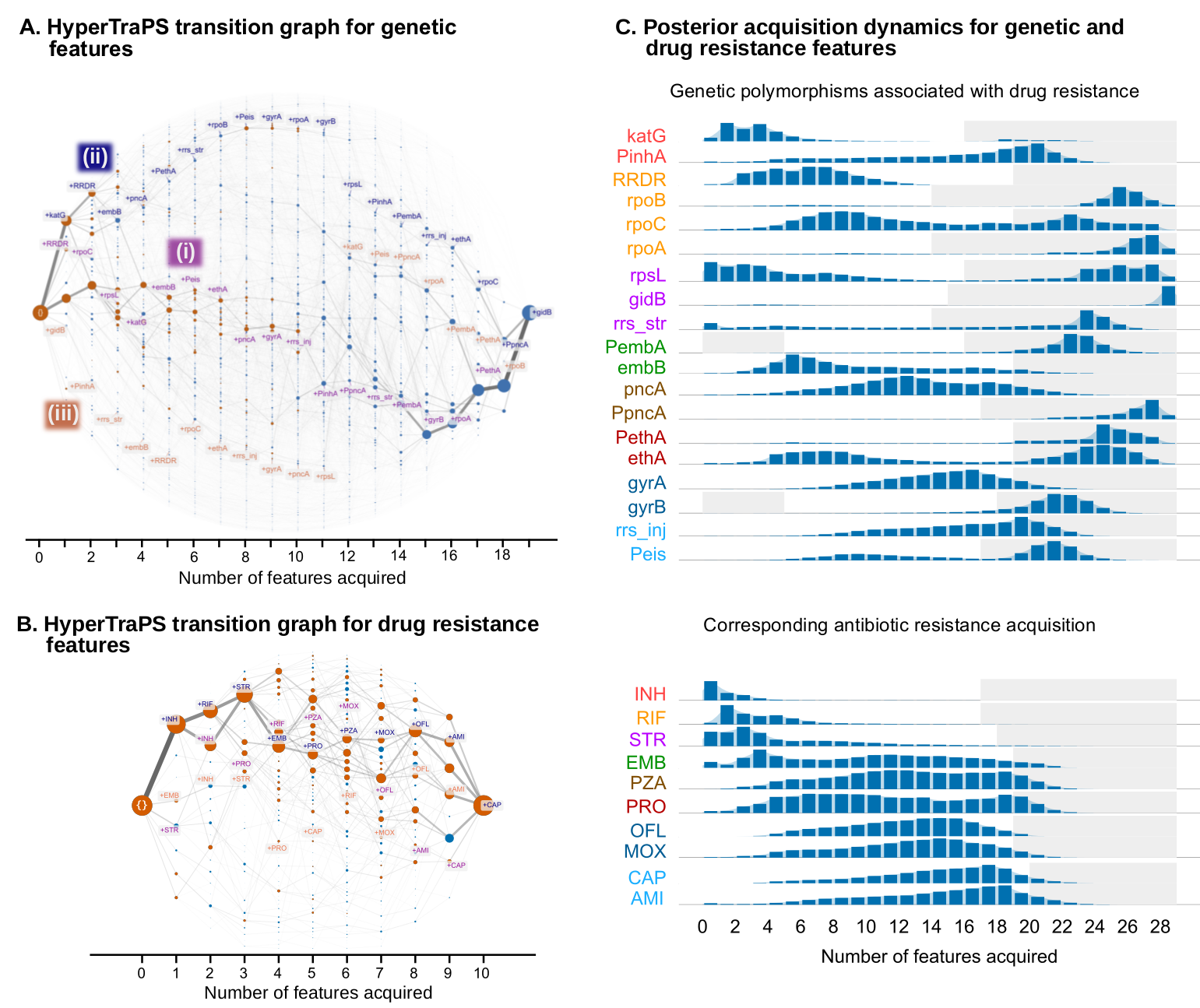}
    \caption{\textbf{HyperTraPS inference for multidrug resistance in tuberculosis identifies different pathway structures linking genetic and drug resistance features.} \textbf{(A)} Dynamic pathways of the acquisition of genetic features leading to drug resistance, inferred from the dataset of $L=19$ genetic sites across $395$ phylogenetically related isolates. Multiple pathways through the genetic space associated with drug resistance are highlighted by regions of different density. Three distinct classes of pathway (i)-(iii) are highlighted and discussed in the text.
\textbf{(B)} Dynamic pathways of the acquisition of resistance to specific drugs, inferred from the dataset of $L=10$ antibiotics across $395$ phylogenetically related isolates. \textbf{(C)} Posterior orderings of genetic and drug resistance features. Rows are colour-coded to link known genetic polymorphisms with the specific drug to which they confer resistance. The genetic sites occupy the first 19 rows, followed by the five `first-line' drugs with the five `second-line' drugs in the last five rows. Density in the grey regions corresponds to acquisitions that do not directly affect the likelihood, as features are not observed to be acquired in these regions in the dataset.}
  \label{fig:tb}
\end{figure*}

We next asked whether our HyperTraPS approach could efficiently characterise dynamics in a system where observations are phylogenetically related. To this end, we consider the case of pathways of genetic polymorphisms that underpin drug-resistant tuberculosis isolates reported in \citet{Casali2014}. In this study, the authors considered the sequences of 1000 drug-resistant tuberculosis isolates from Samara in Russia. The data consists of presence/absence markers of polymorphisms at 16 key genes/promoter regions that confer drug-resistance, as well as mutations in three RNA polymerase genes, and susceptibility or resistance to ten drugs for each of 395 isolates. These observed isolates are linked by a phylogeny, which Casali \emph{et al.} constructed from genome-wide information (importantly, consisting of a much wider set of genomic regions than just those involved in drug resistance). \pr{As in Fig. \ref{fig:methods}, the source data then consists of the states on the leaves of a phylogeny and a phylogenetic structure that is previously, and essentially independently, constructed.}

We assume that mutations are sufficiently rare such that convergent evolution is not a leading-order dynamic process between descendant and parent nodes in the phylogeny. With this assumption, we work backwards through the phylogeny parsimoniously to estimate unobserved parent states. From these estimates, we can reconstruct the transitions from parent nodes to \Green{descendant} nodes on the phylogeny. These transitions then form the observations used by the HyperTraPS platform (Fig. \ref{fig:methods}). \pr{In \sifigfifteen, we characterise the effects of this phylogeny on our posteriors, showing that its detailed structure has only limited quantitative influence on the general pathways we identify.}

\Blueb{In Fig. \ref{fig:tb}A we show the inferred hypercubic transition graph for the dataset with $L=19$ genetic sites alone, highlighting the genetic pathways by which polymorphisms may be acquired.
Once more, a collection of previously unreported dynamic pathways are immediately observed, illustrated by the differential density of edges in different regions of the plot. In contrast to the large number of highly focussed paths inferred from the ovarian cancer data, this transition graph demonstrates a smaller number of looser -- but still distinct in structure -- paths across the hypercube, each with a `cloud' of variability indicating some flexibility in specific orderings within these pathways. We highlight this diversity with three specific pathways: (i) a central common pathway with a rifamycin resistance mutation {\em RRDR} is acquired first along with a fitness compensatory mutation {\em rpoC} second; (ii) an alternative path where no genetic correlates of streptomycin resistance (usually acquired early) are acquired until the sixth acquisition; (iii) a third pathway where the most common polymorphism {\em katG} is acquired late (the twelfth acquisition).}

\Blueb{In Fig. \ref{fig:tb}B we show the inferred hypercubic transition graph for the dataset labelling resistance or susceptibility to each of the $L=10$ antibiotics. The corresponding transition graph reports phenotypic pathways, existing in parallel with the genetic pathways in Fig. \ref{fig:tb}A. Notably, these phenotypic pathways are more canalised than the inferred genetic pathways. Resistance to `first-line' drugs -- those that are first used in treatment -- dominate the initial dynamics, with comparatively little variation in ordering (isoniazid-rifamycin-streptomycin-ethambuol being a common pathway). There is more variation in dynamics of resistance acquisition to the remaining `second-line' drugs, with acquisition subsequently progressing through several different pathways.}

Fig.~\ref{fig:tb}C shows the acquisition ordering plot for the combined genetic and phenotypic state of strains. Competition between different genetic pathways is reflected in the multimodality of several polymorphism acquisition distributions. Notably, \emph{katG}, \emph{rpoC}, and \emph{rpsL} display ordering bimodality, evidencing several different pathways in which these features may be acquired early or late but not at intermediate orderings. This structural flexibility gives rise to the separated pathways discussed above for Fig.~\ref{fig:tb}A. The flexibility in genetic pathways corresponding to first-line drug resistance (for example, \emph{katG}-\emph{PinhA} and \emph{PinhA}-\emph{katG}, both leading to isoniazid (\emph{INH}) resistance) provides a potential explanation for the early acquisition of resistance to these drugs.

Consistent with the more canalised phenotypic pathways in Fig.~\ref{fig:tb}B, there is less multimodality in the ordering distributions of drug resistance features. Resistance to the first line drugs typically occurs before the second line drugs with a more precise order, likely indicative of the more widespread and increased time that tuberculosis has been treated with first line drugs. The ordering in which second line drug resistance is acquired is more broad, agreeing with the flexible phenotypic pathways seen above. Further, despite some heterogeneity, notable dynamic correlations may be observed between drugs and their known genetic correlates. The gene {\em katG} and drug isoniazid ({\em INH}), {\em rpsL} and streptomycin ({\em STR}), {\em embB} and ethambuol ({\em EMB}), illustrate clear examples of such links, providing a predictive and probabilistic connection between the dynamic acquisition of polymorphisms and the acquisition of specific drug resistance phenotypes.

Taken together, this dynamic pathway inference yields several new insights into the structure and variability of the evolutionary trajectories by which drug resistance is acquired. We discuss some specific evolutionary implications in Methos Appendix, and compare with outputs of the approach of \citet{Bollback2006} in Methos Appendix (\sifigsixteen). Broadly, the joint polymorphism and drug resistance dynamics results suggest a consistent, convergent dynamic adaptation to first-line drugs, followed by more heterogeneity in the adaptation to second-line drugs. This convergence in first-line adaptation is likely facilitated, at least in part, by the flexible genetic pathways corresponding to these phenotypes (as found in other convergent evolution examples \cite{Williams2013}). These separate pathways (for example, those involving early vs late polymorphisms in \emph{katG}, \emph{rpoC}, or \emph{rpsL}) are naturally distinguishable from the structures in Fig. \ref{fig:tb}A and multimodality in Fig. \ref{fig:tb}C. The HyperTraPS posteriors further provide a predictive framework which in future can be applied, for example, to predict the next likely drug resistance acquisitions given that a strain is in a particular state.

\section{Discussion}\label{section:discussion}


\begin{table*}[!t]
\centering
\includegraphics[width=\textwidth]{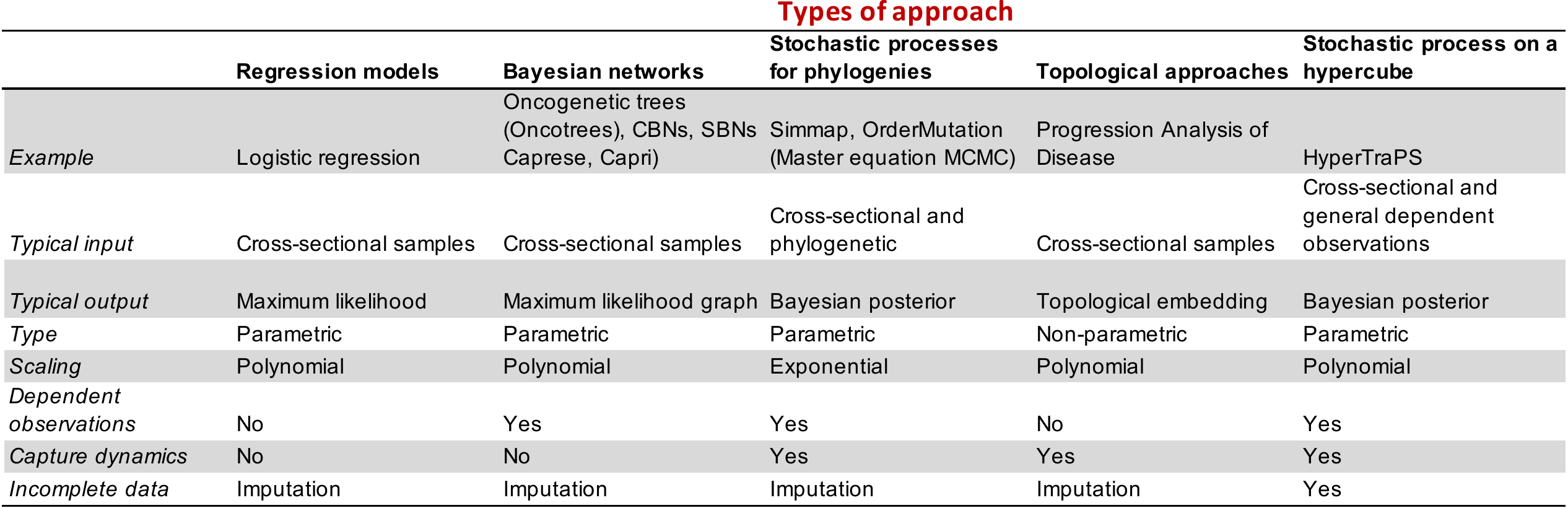}
\caption{\Sam{{\bf Comparison of HyperTraPS with other methods for inference from state space observations. We consider some of the key properties that HyperTraPS introduces.} The following abbreviations are used:
Suppes-Bayes Network (SBN), Conjunctive Bayes Network (CBN) and Markov chain Monte Carlo (MCMC).
}}
\label{table:method_comparison}

\end{table*}


We have introduced \Red{a powerful and highly generalisable statistical platform} for \Green{inferring} probabilistic, coupled dynamics from samples in a binary state space. The generality of this question is illustrated by the diversity of existing approaches that have some bearing on the corresponding inference problem. Table \ref{table:method_comparison} illustrates several broad classes of these approaches, including regression models, Bayesian network models, stochastic processes on phylogenies, topological approaches and finite state space models (HyperTraPS).

Regression models are applied widely across the statistical and biomedical community, but are usually reliant on a linear underlying model and do not attempt to capture dynamics in which variables evolve. Additionally, they require a clear dichotomy between predictors and response variables to be imposed {\em a priori}, when such a distinction may not be appropriate, especially from the perspective of the inference of pathways. \Sam{Bayesian networks provide a common platform for the relationships between features to be learned, with two examples being Conjunctive Bayes Networks \citep{Beerenwinkel2007} and Suppes-Bayes networks \citep{Loohuis2014}. These are commonly used in oncogenetic inference problems\Green{, and have proved successful at unpicking causal relationships between features. We have shown that HyperTraPS aligns with the outputs of these approaches in simple cases. In more general settings, the stochastic model underlying HyperTraPS has the potential to reveal more detailed dynamic structure, including the identification of competing stochastic pathways, complex sets of interactions between coupled traits, and the quantification of uncertainty in the pathway structures that are revealed.}}

Dimensionality reduction approaches have been considered for finding representations of temporal dynamics from samples. Such methods are powerful and have been applied to vast data collected in whole genome single cell RNA experiments \citep{Campbell2016} and also to disease \citep{Nicolau2011}. While highly flexible, these approaches often rely on specific assumptions about the quantitative details of the dimensionality reduction, leading to variability from method to method, and have yet to be considered in detail for finite space state models like the presence/absence structures we consider here.

Modelling trait evolution on phylogenies is the closest group of models to which HyperTraPS is related, and typically requires computation of master equation rate matrices that do not place restrictions on the transitions that may occur in the state space \citep{Bollback2006, OMeara2012}. By embedding transitions on a hypercubic graph, HyperTraPS has the ability to handle orders of magnitude more features without noticeable loss of generality (simultaneous transitions are represented as equally weighted, temporally adjacent, transitions). Additionally, these methods are designed specifically for phylogenies, while HyperTraPS has applicability to generic sample dependency.

The HyperTraPS framework presented here has several advantages: (a) its polynomial scaling allows it to deal with large (many observations and many traits) datasets; (b) the regularisation processes we outline allow it not only to \Green{reveal and deal with} arbitrary coupling between traits, but to select good and statistically significant parametric representations of these couplings to yield sparse models \Green{(thus applying Occam's razor)}; (c) it yields general and readily interpretable predictions; (d) it simultaneously provides \Green{inferred pathway structure, mechanistic insight, and uncertainty quantification}; (e) the ability to include prior information about pathway structure when existing knowledge about biological mechanisms forbids, disfavours, or enhances the probability associated with particular transitions. Despite these advantages, there are of course some limitations to the platform's capabilities. Incomplete data currently provides a challenge for inference with HyperTraPS. There is nothing in principle preventing \emph{hypercubic inference} with incomplete data: unbiased random walks can be simulated on a hypercube and their ability to recapitulate observations can be computed. Indeed, HyperTraPS can be applied in the case of uncertain \emph{end points} of observed transitions (representing an advantage over existing methods). However, the sampling algorithm that allows HyperTraPS' efficient sampling of high-dimensional spaces currently does not translate to incompletely described \emph{start points} of observed transitions, requiring future work is needed for further generalisations. Further, our approach for regularisation, while successfully implemented above, relies on an imperfect greedy algorithm and on the subjective use of the Akaike Information Criterion (AIC) for finding such sparser models. A multitude of methods are available for performing model selection within a full Bayesian setting \citep{OHara2009, Murphy2012} and exploration of alternative approaches for exploration of mappings from $W \rightarrow \pi$ and regularisation of HyperTraPS models is an important future avenue of research.


\Red{Our platform occupies the under-explored intersection between methods for inferring dynamics from uncoupled and/or longitudinal observations (as in cancer progression) and from phylogenetically linked observations (as in evolutionary inference). We have shown that HyperTraPS has a unique power to dissect multiple competitive dynamic pathways \Green{(yielding new insight in two biomedical case studies)}, and demonstrated how the processes of regularisation can be used to identify the best model structures for a given scientific setting. We underline that HyperTraPS requires no domain-specific knowledge, but can readily include such knowledge in the form of priors and in posterior interpretation. The platform is therefore ideal for contexts where mechanistic insight and modelling are less developed, and hence may also find valuable use in the wide range of progressive diseases that are less studied than cancer. We anticipate that this flexibility, and the abilities of HyperTraPS to naturally quantify uncertainty and form probabilistic predictions about future behaviours, will be of use across biomedical, evolutionary, and other scientific disciplines as volumes of available data continue to increase.}

\section{Lead Contact and Materials Availability}
Requests for further information and resources should be directed to and will be fulfilled by the lead contact, Iain Johnston (iain.johnston@uib.no). This study did not generate new unique reagents.

\section{Data and Software Availability}
\Sam{All computational work was performed with custom-written software in C++ and Python. The code for the HyperTraPS package is freely available \pr{at https://github.com/sgreenbury/HyperTraPS} (DOI 10.5281/zenodo.3478290) and usable under the creative commons licence.}

\section{Acknowledgements}
All authors acknowledge support from EPSRC grant EP/N014529/1. IGJ additionally acknowledges support from a Turing Fellowship from the Alan Turing Institute.

\section{Author Contributions}
Study concept and design: SFG, MB, IGJ; Development of source code: SFG, IGJ; Analysis and interpretation of data: SFG, MB, IGJ; Writing and revision of the manuscript: SFG, MB, IGJ; Study supervision: MB, IGJ.

\section{Declaration of interests}
The authors declare that there is no conflict of interest.

\bibliographystyle{apalike}
\bibliography{hypertraps_methods}

\newpage
\beginstar

\section*{Methos Appendix}
\section*{HyperTraPS pipeline}

In \sifigone, we provide a diagrammatic overview of the HyperTraPS pipeline. The different elements are described below. \Blue{As described in Fig. \ref{fig:methods}, the first step is to convert cross-sectional, longitudinal, or phylogenetically linked observations to a set of transitions, which we will represent as $D = \{s_i, t_i\}$, where $s_i$ is the $i$th source state and $t_i$ the $i$th target state, and there are $n_D$ observations in total.}

\section*{Bayesian framework and likelihood of transition dataset}
As introduced in \citet{Johnston2016}, we choose a Bayesian framework for inferring parameters for the set of edge weights $W$ on the \Green{hypercubic transition graph} that explain the data \Green{$D$}.

As such we are concerned with drawing samples from the posterior:
\Blue{$$
P(W | D) = \frac{P(D | W)}{\int P(D | W)P(W)dW}P(W)
$$}
which is proportional to the product of our prior probability density \Green{$P(W)$} on edge parameterisations and the likelihood \Blue{$\mathcal{L}(W | D) = P(D | W)$}, such that we have \Blue{$P(W | D) \propto \mathcal{L}(W | D)P(W)$}. Throughout this work we choose a uniform prior distribution on $P(W)$ and therefore only need to consider the calculation of \Blue{$\mathcal{L}(W | D)$} in order to derive samples from the posterior probability distribution.

From this transition set, we can decompose the likelihood into the following form (regardless of whether the source data was cross-sectional, longitudinal, or phylogenetically coupled \citet{Johnston2016}):
\Blue{$$
\mathcal{L}(W | D) = \prod_{i=1}^{n_D} P_\text{observe} (s_i \rightarrow t_i)
$$}
where $n_D$ is the size of the transition dataset. $P_\text{observe}$, the probability of observing such a transition requires \Green{a signal to be emitted by our system at both the source and target states}, with the system having reached the source state and then made the transition to the target state via any possible walk on the hypercube. Therefore, the probability of observing such a transition can be written as:
$$
P_\text{observe} (s_i \rightarrow t_i) = P_\text{emit}(s_i, t_i) P_\text{reach}(s_i | W) P(t_i | s_i, W)
$$

\Blueb{We assume that signal emission in a given state is a random process that independent of the state. Given the term $P_\text{emit}(s_i, t_i)$ is also independent of $W$, and that we deal only with complete data here, $P_\text{emit}$ yields a constant multiplicative factor which can be ignored in the inference process. In~\citet{Johnston2016}, it is shown that the remaining log-likelihood can be written as:}

\begin{equation}\label{eq:loglik}
\Blue{\log \mathcal{L}(W | D) = \sum_{i=1}^{n_D} \log P(t_i | s_i, W) := l(W | D)}
\end{equation}
where the only computation required is the probability of making the transition to $t_i$ from $s_i$ for a given parameterisation of $W$. 

In order to calculate $P(t_i | s_i, W)$, a sum over all possible paths between $s_i$ and $t_i$ is required. Given that the number of paths between $s_i$ and $t_i$ scales as the factorial of the Hamming distance, the problem of deriving the rate matrix becomes intractable for systems of dimensions \pr{around $L \gtrsim 10$}. Instead we tackle the problem by way of performing biased random walks restricted to pathways that end in $t_i$. This method of sampling was introduced in \citet{Johnston2016} and allows systems with more features to be considered than previously has been the case. This HyperTraPS algorithm that forms the key part of the HyperTraPS framework is captured in Algorithm \ref{alg:htps_complete}.

\begin{algorithm}
\DontPrintSemicolon
\KwData{$D^\text{transitions} = \{s_i \rightarrow t_i \}^{n_D}_{i=1}$}
\KwResult{Estimate of $P(D^\text{transitions} | W)$}
\Begin{
	\For{$(s \rightarrow t) \in D^\text{transitions}$}{
		$s_c \leftarrow s$\;
		Initialise $N_h$ trajectories starting at state $s$\;
		\For{$i \in N_h$}{
			$s_c \leftarrow  s$\;
			$\alpha_i \leftarrow 1$\;
			\While{$t$-compatible move possible for $s_c$} {
				Calculate the probability of making a $t$-compatible move, record as $\alpha'_i$\;
				$\alpha_i \leftarrow \alpha_i \alpha'_i$\;
				Choose a $t$-compatible move at random in proportion to its transition probability\;
				Make move and update $s_c$ accordingly\;
			}	
		}
		$\hat{P}(s \rightarrow t) = N^{-1}_h \sum_i \alpha_i$\;
		$P(D^\text{transitions} | W) \leftarrow P(D^\text{transitions} | W) + \hat{P}(s \rightarrow t | W)$\;
	}
}
\caption{{\bf HyperTraPS algorithm for complete data}: Hypercubic Transition Path Sampling was first introduced by \citet{Johnston2016} to sample random walks on a hypercube across a restricted set of compatible states between a source and target state.}
\label{alg:htps_complete}
\end{algorithm}

\section*{Tractable parameterisations of hypercube}
The transition graph linking states with $L$ features has $L2^{L-1}$ edges that we aim to parameterise. As $L$ grows, we require a way of reducing this number of parameters $k$ without compromising our ability to describe the dynamics of a system. Shrinkage and model selection tools may be used to achieve this reduction: we explore a simple approach for this process later. However, given the \Red{potentially} \Blue{large} number of parameters in the default model, we also consider methods to reduce parameter space before the inference process. 

One intuitive approach is based around considering the factors that may influence a given transition. The full parameterisation allows independent rates between any two states. In this picture, the probability $P(i)$ of acquiring the $i$th trait can take arbitrary and independent values for every possible combination of the other $L-1$ traits. As an alternative, we can restrict the dependence of $P(i)$ on the \emph{coupling} of other trait patterns. For example, if we assume that each of the $L-1$ other traits influence $P(i)$ independently (no synergistic interactions), we need only $L^2$ parameters: a `basal rate' of acquisition for each trait $i$, and the amount by which this basal rate is modified by the presence of trait $j \not= i$. This reduction is analogous, for example, to Generalised Linear Models where response variables can be considered a function of independent variables and interaction terms between the independent variables, neglecting higher order interaction terms.

From this perspective a hierarchy of models may be constructed (\sifigtwo). For the `zero order' model every feature has equal probability of acquisition ($k=0$ parameters). In the `first order' model, every feature has an independent acquisition probability ($k=L$ parameters). In the `second order' model, every feature's \pr{basal} acquisition probability \pr{is independently modulated by} the presence of each other feature \pr{($k=(1 + (L-1))\times L = L^2$ parameters)}. Higher order models, including the full $L 2^{L-1}$ set can be envisaged, introducing more complex interactions between the co-occurrence of features.

\pr{To illustrate these parameterisations, consider the weight $w_{s \rightarrow t}$ of the edge from state $s$ to state $t$. These edge weights are nonzero only for pairs ${s,t}$ where $t$ differs from $s$ by the acquisition of exactly one feature, with a hypercubic network remaining. Then, for the zero-order model, every edge in the hypercube is equally weighted, and we can set this weighting to unity, $w_{s \rightarrow t} = 1$. For the first-order model, the weight of an edge is completely specified by the feature that the edge corresponds to acquiring, $w_{s \rightarrow t} = p_i$, where $i$ is the feature that distinguishes $s$ from $t$. The first-order parameterisation is thus described by the vector $\mathbf{p}$ with $L$ elements, one for each feature. For the second-order model:}

\pr{\begin{equation}
  w_{s \rightarrow t} = p_{ii} \prod_{j \not= i} q(s, j, i), \label{secondorder}
\end{equation}}

\pr{where $i$ is the feature that distinguishes $s$ from $t$, $s_j$ is the presence/absence of feature $j$ in state $s$, and $q(s, j, i) = 1$ if $s_j = 0$ and $p_{ji}$ otherwise. The second-order parameterisation is thus described by the matrix $p$ with $L^2$ elements, where diagonal element $p_{ii}$ gives the `basal' rate associated with feature $i$, and off-diagonal elements $p_{ji}$ give the influence that the presence of feature $j$ in source state $s$ has on this basal rate.}

\pr{For numerical convenience, we implement Eqn. \ref{secondorder} via a logarithmic transformation, such that $p_{ij} = \ln \pi_{ij}$, and work with $\pi_{ij}$ as the parameterisation of the model. We will use $\pi$ generally to refer to edge weight parameters in the inference process.}
  

\section*{Monte Carlo sampling methods}
The complexity of the inference problem challenges analytic or uniform sampling approaches to compute Eq. \eqref{eq:loglik}. Instead, we employ Markov Chain Monte Carlo (MCMC) in order to generate samples from the posterior on edge weights $W$. As the HyperTraPS algorithm generates \Blue{an estimate of the likelihood (with the same expected value as the exact likelihood), this is in fact a pseudo-marginal MCMC sampler} which has been shown to yield the same stationarity properties as if it were exact \citep{Andrieu2009}. 

Previous approaches \Red{for specific scientific questions} \citep{Williams2013, Johnston2016} found this pseudo-marginal MCMC sampler \Red{to demonstrate} good mixing. However, there are cases where this simple approach produces poor mixing, \Purple{specifically when the Hamming distance between a source and target state becomes large}. This is because Algorithm \ref{alg:htps_complete} generates an estimate of the likelihood with increased variance around its exact value due to the greater number of acquisitions made during path sampling. This can lead to poor mixing of sampler chains, if the sampler draws a high value for the likelihood estimate which subsequent random draws have a high probability of having a lower likelihood for the same parameterisation. This occurs when the variance of the total log-likelihood has a variance with magnitude greater than unity \citep{Sherlock2015}.

To address this issue and generalise to more diverse datasets, we embedded HyperTraPS within an auxiliary pseudo-marginal MCMC algorithm (APM MCMC), which also satisfies the same convergence properties as MCMC \citep{Murray2015}. By making the likelihood a joint density $l(\pi, u)$ over the parameters of the model and also the random variable from which our estimate is drawn, alternate Metropolis-Hastings steps can be performed by keeping $\pi$ and $u$ alternately fixed during the proposed update to the chain. For HyperTraPS, a new proposal for the random variable $u$ is a new set of random trajectories across the hypercube over which each observations's likelihood is estimated. We make use of this scheme throughout this work, as little computational overhead is introduced, and mixing times are dramatically improved.

As discussed, we have assumed a uniform prior for the parameterisations of the hypercube. For our choice of mapping $\pi$, this means we choose $P(\pi) \in U(-m,m)$ where $m=10$ and $m=20$ are used in this work to cover several orders of magnitude of relative size across the inferred parameters.

We begin MC sampling runs with the parsimonious initial condition $\pi=\mathbf{0}$. This is equivalent to the zero order model where there is no directionality pre-supposed. This facilitates the avoidance of local traps in the parameter landscape while remaining agnostic in introducing directionality into the inferred parameterisations for a particular dataset. \Blueb{A burn-in period occurs before expected convergence of an MCMC chain. For each of datasets in the main text, over $10^6$ iterations are performed along the chain, ensuring samples are used only when convergence is apparent. We consider convergence to be reached when the chain shows stability in average likelihood for a sustained period with the ratio of accepted parameterisations that yield increased or decreased likelihoods to be in equal proportion.}

\section*{Simulated walks to illustrate order of acquisition}\label{section:simulated-walks}
The inference process above yields inferred posterior distributions on the hypercubic edge weights $W$. We can query these posteriors in a number of ways to gain descriptive and predictive information about the mechanisms generating observed states. First, we produce a parsimonious and intuitive representation of the dynamic pathways supported by the inferred posteriors. \Red{Here, we} simulate an ensemble of random walkers generating complete trajectories on hypercubes with sets of transition probabilities sampled from the inferred posterior. This ensemble reflects the likely dynamic pathways supported by the dynamic transition model after parameterisation. We simulate an ensemble of random walks in two ways: {\em Walk Simulation 1 (WS1)}, with walkers that run from $\{0\}^L$ to $\{1\}^L$ where a feature is acquired at every time step and {\em Walk Simulation 2 (WS2)} only simulates trajectories corresponding to transitions observed in the dataset. In each case, we record \Blue{every transition between states allowing the construction of a weighted directed graph of all states and transitions encountered. From this graph,} the frequency $f_{ij}$ with which feature $i$ is gained at step $j$. 

\section*{Graph embedding and visualisation for dynamic acquisition on the hypercube}
With each simulated random walk, $L$ transitions occur between states on the hypercube. Across a large sample of random walks, we define this set of states as $\mathcal{S} = \{s_i\}$ and we can represent the number of transitions between any two states by a directed, weighted graph with adjacency matrix $a_{ij}$.

In order to visualise this graph to reveal characteristic progressions across the hypercube resulting from a given parameterisation, we use a custom embedding to project the high-dimensional graph into two dimensions. First, we project the hypercube on to the surface of a sphere and optimise the projection by making the following choices:
\begin{itemize}
\item Every state is given the same radial coordinate, $r=1$.
\item The number of features acquired in the dataset is a measure of the how far the state is along the progression from $0^L$ to $1^L$. Therefore, for every state $\mathcal{S}$, we count the number of acquired features ($n$ out of $L$) and assign a polar angle $\theta$ such that $\sin \theta = n/L$.
\item The azimuthal angle $\phi$ on the interval $0\leq \phi \leq \pi$ is assigned by considering the {\em mean angle of the states from all incoming edges}, therefore attempting to maximise the potential spread of the most common distinct paths across the hemisphere. A final assumption involves choosing all states with a single acquisition ($L$ states) to be uniformly spread on the cosine of the interval $[0,\pi]$.
\end{itemize}

With the embedding, the plot of the adjacency matrix $a_{ij}$ is augmented by choosing node sizes and edge widths in proportion to the number of times the state and the transition are respectively encountered by the ensemble of random walks. \Sam{Three examples of plots generated from this embedding with parameterisations of the hypercube are shown in Fig. \ref{fig:cs12}B(i)-(iii)}, illustrating the ability to display different underlying progressions inferred with HyperTraPS.

\Sam{In presenting the embedding, we adjust the graphical depiction to highlight the features of the graph in the following way:}
\begin{itemize}
\item Vertex area is in proportion to the number of times the vertex is visited by WS1 simulated random walks.
\item Edge widths and opacity are in proportion to the number of times the transition between states is made with a random walker under WS1.
\item States encountered are coloured blue if $s \notin D^\text{transitions}$ and orange if $s \in D^\text{transitions}$
\item To highlight representative paths across the hypercube, we employ a labelling scheme as follows.
As walkers start from the empty ``$\{\}$'' state ($\{0\}^L$), we can consider the addition of single features as each edge is traversed. In the plots, we use a greedy mechanism for determining which edges to label. Starting from $0^L$, we take the most probable outgoing edge at each vertex encountered and label the feature acquired across that edge at the resulting vertex until the $1^L$ state is reached, giving is the first greedy path. The following $n$ greedy paths make use of the same approach but disregard any previously labelled edges, taking the next most probable available. We use the approach to clearly identify the left-right and right-left paths in Fig. \ref{fig:cs12}B(i)-(iii).
\item Finally, an optional transform to remove vertex overlap may be applied to remove overlap of vertices with a given number of features, while retaining the relative area of each vertex that is determined by the number of times the vertex is encountered.
\end{itemize}

\section*{Probabilistic feature graph representation}
Using either WS1 or WS2, the set of states encountered may be considered as a directed weighted acyclic graph through {\em sample space}/{\em state space}, due to the irreversible acquisition of features. As paths through state space involve the acquisition of a feature with each incoming and outgoing edge, a different representative graph may also be constructed relating the observed consecutive feature acquisitions producing a graph in {\em feature space}.

To this end, we consider the ensemble of observed $P(Y_{out}, X_{in}; s)$ derived from a set of simulated walks across sample space, which gives the probability that feature $Y$ is acquired leaving state $s$, with feature $X$ having been acquired to reach state $s$.
An average joint relationship can then be written as the following:
$$P(Y_{out}, X_{in}) = \sum_s P(Y_{out}, X_{in}; s)P(s)$$
where $P(s)$ is the proportion of times state $s$ is encountered.
$P(Y_{out}, X_{in})$ gives the edge weight between X and Y for the probabilistic feature graphs in this article.

\section*{Regularisation}
We previously discussed approaches to reduce the parameter space of the HyperTraPS model while retaining dynamic information. We can {\em a priori} also employ model reduction approaches to identify supported parameter structures given a particular dataset. This regularisation helps identify more interpretable, parsimonious models and to guard against over-fitting.

\Green{One approach to model selection would be a fully Bayesian exploration of the joint space of model structures and parameters. However, the combinatorial explosion of search space with $L$ currently makes this approach unfeasible for all but the simplest systems. Instead, we sacrifice a full exploration of this complicated space in favour of a tractable but principled approach to balance the reduction of model complexity against the ability to fit the data. This} illustrative metric can indicate the amount of redundancy present in the parameterised $\pi$ that can be removed in order to reduce the potential for over-fitting. To this end, we introduce a cost function to penalise the log-likelihood and then perform a algorithmic search to optimise this function.

We note that the number of parameters $k$ required to adequately describe a given dynamic system is deeply related to the mechanisms underlying that system. \Blueb{If features are acquired independently, the first order model with $L$ parameters should be sufficient to capture the dynamics (as seen in Section \ref{section:csd} of the main text for dataset $D_1$), and the features may be completely ordered for the average trajectory.} If a higher order model with more parameters is required, it suggests that interactions exist between features, such that one feature may influence the acquisition propensity of another. Identifying the sparsest model that can account for observations therefore also reveals mechanistic insight into the system.

For simplicity, we use the Akaike Information Criterion (AIC) \citep{Murphy2012} to introduce sparsity. The AIC score for a model can be written as:
$$
\text{AIC} = 2(k - \hat{l})
$$
where $k$ are the number of parameters in the model, and $\hat{l}$ is the maximum log-likelihood. The score comprises the log likelihood and a penalty for lack of sparsity, in this case, the number of non-zero elements included in the maximum likelihood parameterisation $\pi$. Other options for regularization scoring include the Bayesian Information Criterion (BIC), but we refrain from exploring different metrics here, focussing firstly on illustrating how such regularisation can be performed within the HyperTraPS framework. A more general model selection approach will be the subject of future work.

\Blue{To find parameterisations that optimise the AIC, we take a {\em greedy backward selection} approach \citep{Murphy2012} to reduce the number of parameters $k$ for a given model type. The process can be applied to both the first- and second- order models.
An issue with such a greedy approach is that each single greedy backward step is unable to account for interactions between multiple parameters that lead to lower scores. Therefore, given a set of potentially distinct approximately maximum likelihood parameterisations, different backward selection processes from different starting maximum likelihood models may yield different minimum AIC scores for a given value of $k$. In an attempt, to bypass this problem, we take an ensemble of the top 100 maximum likelihood parameterisations from an MCMC sampling procedure \Blueb{(top 1000 for the ovarian cancer datsets)} and perform the greedy backward selection process to each one. Across the ensemble, for a given parameter number $k$, we take the minimum AIC score as a proxy for the minimum model at this level of parameterisation. The global minimum with respect to AIC is taken as the {\em first order regularised} or {\em second order regularised} model for the a first order and second order starting point respectively. The regularised models are then taken used in the subsequent section to perform model validation.}

\Sam{In Fig. \ref{fig:cs12}D(i)-(ii), we show the regularisation process described above for the minimum of the ensemble at each value of $k$ for the two synthetic datasets $D_1$ and $D_2$ and, later in Methos Appendix, the process for a third synthetic dataset and the ovarian and tuberculosis datasets respectively.}

\section*{Validation}
\Blue{Importantly, the inferred parameterisations from our approach can be used to predict future behaviour for a given state. We have described two procedures for generating parameterisations: sampling from the full posterior for a given model (first- or second- order) or regularised parameterisations constructed by the procedure in the previous section. In this section, we perform model validation through using the regularised parameterisations in order to identify the strength of evidence for the first- or second- order models. Using the outcome of this procedure, either samples from the full posteriors of the identified model or from the corresponding regularised parameterisation can be used for prediction.}

\Blue{We validate this predictive power through two methods: firstly, through basic model comparison between the regularised first- and second- order models; and subsequently, by calculating the likelihood of observing data not used in the inference part of the method as a proxy for the predictive capability of each model.
As a simple procedure to illustrate this, we split the $D^\text{transitions}$ dataset into two halves: a training dataset $D_\text{train}$ on which samples from the posterior are drawn and model comparisons can be made, and a testing dataset $D_\text{test}$ with which the likelihood can be calculated using samples from the posterior for $D_\text{train}$.}

\Blue{For model comparisons, we choose the zero order model as a null model. For comparisons between the different order models, we find the regularised first- and second- order model for the training dataset and denote this likelihood as $\hat{l}(\pi | D_
\text{train})$. We then perform a likelihood ratio test, using the log-likelihood ratio statistic (LLR):
$$
LLR = 2\hat{l}(\pi^{(j)}_r | D_\text{train}) - 2\hat{l}(\pi^{(0)}_r | D_\text{train})
$$
where $\pi^{(j)}_r$ is regularised $j^\text{th}$ order model. We compare to the $\chi^2$ distribution for the number of non-zero parameters in $\pi$. With regard to the test dataset, we then use HyperTraPS to estimate $\log P(D_\text{test} | \pi^{(j)}_r)$ providing a measure of predictive capability of the $j^\text{th}$ order regularised model. This is an intuitive option for measuring performance as it is not guaranteed that a given transition from $s\rightarrow t$ should end at $t$ -- there may be multiple pathways. Therefore, the overall largest likelihood ($\log P(D_\text{test} | \pi^{(j)}_r)$) across competing $j$ models for the test dataset will be monotonically related with better parameterisations.}

\section*{\pr{Testing and validating HyperTraPS with differing data structures, volumes, and priors}}\label{section:d_post_rev_1}

\pr{In this section we investigate HyperTraPS' capacity to learn pathway structures by varying several features of the synthetic datasets used in the main text. Fig. \ref{mtnewfig} in the main text provides central aspects of this investigation; \sifigthree involves different relatednesses of observations; \sifigfive provides posteriors for the investigation of different numbers of competing pathways; \sifigsix provides probabilistic feature graphs for the use of prior information.}

\pr{Both quantitative and structural prior information about models can be included in HyperTraPS. Quantitative information (for example, the acquisition of one feature scaling the acquisition probability of another) can readily be included through applying an appropriate prior distribution on the corresponding element of the transition matrix. Simple structural information, such as forbidding one transition before another, can also readily be captured by setting priors on the corresponding parameters.}

\pr{Prior information can also be incorporated where an underlying tree structure of precedence between features is known. We denote this the {\em prior tree}. In order to incorporate such information, we wish to avoid parameterisations of $\pi$ that would violate the ordering described within a prior tree. In order to prohibit transitions in the second-order $L^2$ parameterisation system, for a given edge in the prior tree $a \rightarrow b$, we enforce a prior with low basal probabilities for the acquisition of $b$ (in proportion to the depth of $b$ from the root in the prior tree). That is, spontaneous acquisition of states below the root in the prior tree is enforced to be highly unlikely. We then enforce a prior with off-diagonal elements so that the acquisition of $a$ compensates this low basal probability on the acquisition of $b$. Hence, other features aside from $a$ remain unable to affect the acquisition of $b$, but once the precedent feature $a$ is acquired, then $b$ may be acquired.}

\pr{To demonstrate this approach, consider a prior tree with edges: $R\rightarrow1\rightarrow2; R\rightarrow3$, for $L=3$, where $R$ corresponds to the root of the prior tree. Starting from a uniform prior $U(-m, m)$ on all elements of $\pi$, we enforce three prior requirements. First, $\pi_{22} < \pi_{ii} - \Delta$ for all $i \neq 2$ (enforcing low basal acquisition for feature 2). Second, $\pi_{12} \geq \Delta$ (allowing the acquisition of feature 1 to `rescue' this low basal rate). Third, $\pi_{i2} = 0$ for all $i \neq 1$ (allowing no other acquisitions to `rescue' the low basal rate). In this way, we ensure an acquisition probability of 2 prior to 1 or 3 is suppressed by a factor of $e^{\Delta}$. In practise we have used $\Delta = dm / l$, where $d$ is the depth of a feature in the prior tree, $m$ as above is the range of the original uniform prior, and $l$ is the maximum depth of the prior tree.}

\section*{Additional synthetic cross-sectional dataset}\label{section:d_cs_3}

In this section, we illustrate the inference, regularisation and predictions with a third cross-sectional dataset $D_3$. This dataset can be considered a composite of previous synthetic sets $D_1$ and $D_2$, such that new set $D_3$ is the linear combination $D_3 = 2D_1 + D_2$. In this case, we have a dominant progression underlying the dataset but with a substantial minority contribution from an alternative pathway.

In \sifigseven A, the structure of this additional cross-sectional dataset is depicted. \sifigseven B, C and D indicate that HyperTraPS can infer the two distinct progressions and the proportion with which these progressions occur within the data. For example, in the density plots, feature $i=0$ is acquired three times as frequently as feature $i=7$ in for step $j=0$.

In \sifigeight A and \sifigeight B, we show the results of regularisation and the outputs of the validation methodology: the first order model can be observed to be a better predictor than the null model (it captures the dominant progression) as seen with larger and significant log-likelihoods for the full and training datasets. The second order and regularised second order models perform much better still by having the ability to capture both the dominant and secondary progressions present in the dataset, as illustrated in the validation methodology by the much larger associated likelihoods.

\section*{Alternative interpretation of inferred acquisition orderings (`Walk Simulation 2')}\label{section:ws2}
In Methos Appendix above we introduced a protocol for using samples from the posterior of $\mathcal{L}(\pi | D)$ to illustrate the order in which features are acquired. We denoted this process Walk Simulation 1 (WS1) as simulations from $\{0\}^L$ to $\{1\}^L$ are performed with the feature $i$ acquired at step $j$ being recorded as a proportion $f_{ij}$. As a feature is a always gained in each step, and all features are gained at some stage during this simulation process, the two properties $\sum_k f_{kj} = 1$ and $\sum_k f_{ik} = 1$ both hold. We illustrated the result of this simulation using a histogram for the matrix $f_{ij}$ with kernel density estimates overlaid for each feature.

An alternative simulation protocol is to only simulate trajectories corresponding to transitions that are observed in the dataset. In other words, rather than assuming random walkers proceed from ${0}^L$ to ${1}^L$, we simulate a set of walkers between each pair of source and target states $s_i, t_i$ in the dataset, relaxing the requirement that walkers start at ${0}^L$ and end at ${1}^L$. We denote this process Walk Simulation 2 (WS2). For WS2, we can consider $f_{ij}$ as the probability:
$$f_{ij} \approx P(\text{feature } i \text{ is gained at step } j | s=\{0\}^L \rightarrow t=\{1\}^L)$$
where $s$ is the source state and $t$ is the target state of the set of random walks. Summation over the rows or columns of $f_{ij}$ no longer hold as there is no guarantee in the data that a feature is acquired at a given step $j$ or that every feature $i$ is acquired in each random walk.

\Purple{The main distinction between WS1 and WS2 is the following: \Green{WS1 infers trajectories, informed by data, that start at $\{0\}^L$ and acquire all features to reach $\{1\}^L$. WS2 restricts the inference to the region `covered' by the set of transitions observed in the dataset. Therefore, WS1 provides a readout of a complete process of acquisition (so may be more appropriate for analysis in systems where this is the expected outcome), while WS2 gives a readout of trajectories without extrapolating beyond the limits of observed states} (and may be more appropriate if the walks are not believed to go to completion).}

\Purple{We plot the densities for WS1 and WS2 in \sifignine for two datasets from the main text, synthetic set (ii) and the tuberculosis dataset.
As a result of this different approach, there are three key differences. \Green{First, posterior probabilities are rescaled according to how much a trait is `covered' by observations. This is seen, for example, in feature 1 (and feature 16) in \sifignine A. Here, under WS1, early and late acquisitions of the feature are inferred to be equally likely, as walks are inferred to always run to completion. Under WS2, the number of walks that run to completion is lower (only some observations include `complete' acquisition). The early acquisition mode is then inferred to be more likely, with a balancing probability that the feature is \emph{not} acquired.}}

\Sam{Secondly, with WS1, as the process starts from $\{0\}^L$, for a single random walk, the transitions observed in the dataset are not guaranteed to be reached by random walkers. This means that the overall inferred parameterisations across the entire dataset may not lead to transitions in the dataset being encountered for a finite ensemble of random walks. As a result, the WS1 process does not allow us to directly consider solely the acquisitions between states in the original transition datasets. By exactly considering these transitions, WS2 allows this data to be examined using the parameterisations that have been sampled across the entire dataset allowing for a different type of inference. A clear example of this is seen in \sifignine B for feature {\em PembA} or {\em PethA} that are rarely encountered in the window of acquisition where they are acquired in the dataset, illustrated by the strikingly different distributions for WS1 and WS2.}

Thirdly, there is no density observed in the grey regions for WS2 due to there being no transitions in the dataset \Green{`covering'} these regions, so no transitions performed with WS2 record any density there. In \sifignine B, in application to the tuberculosis dataset, the lack of WS2 density in the grey regions is apparent. In addition, there is clearly observable multimodality in WS2. Multimodality in WS1 is indicative of a feature belonging to multiple progressions that \Green{may include an absence of acquisition if the trajectory does not terminate}. In contrast, multimodality in WS2 is indicative of multiple progressions where multiple orders of acquisition of a given feature \Green{are directly} observed in the data. A striking example is {\em PethA} where in WS1 the predominant visible mode of acquisition is in the grey region towards the end of all possible acquisitions, while in WS2, the acquisition is observed in two distinct regions at step $j=5$ and step $j=10$, suggesting that the transition data contains multiple types of progression where {\em PethA} is acquired. This is also clearly the case for other features such as {\em PembA}, {\em PinhA}, {\em ethA} and {\em RRDR}.

\Purple{We introduced WS2 here as a supplementary form of enquiry of the posteriors that can potentially reveal additional inferences about the underlying progressions from which the data may be derived. In the next section, we look in more detail at the assumptions, types of progressions and the outputs in the plots we have used for the inference in order to motivate intuition further.}

\section*{Implicit assumptions and interpretation of parameterisations}

Here we consider several features of datasets that may be considered challenges to inference with HyperTraPS, and illustrate the corresponding outcomes of our approach:

\begin{itemize}
\item \Green{{\em No structure}: only in the case of independent feature acquisition and identical frequencies will no suggestive progression be found, \Green{in which case the prior distribution (in this article, uniform across all trajectories) will be recovered by the inference process.}} 
\item {\em Samples from complete and partial progressions}: If one or more of the underlying progressions does not correspond to a complete walk across the hypercube, \Green{transition density in unsampled regions will be dictated by extrapolated dynamics or the prior, depending on whether WS1 or WS2 is used.} In \sifigten (i) we illustrate the synthetic dataset (i) for $L=8$ but for a progression that now stops after gaining feature $i=4$. In this case, with no other progressions present in the dataset, we find that the remaining features gained in the grey region do so \Green{with a uniform distribution over remaining orderings (recovering the prior)}. In \sifigeleven (ii) we examine the case where there is a complete right-left path and a partial left-right path (that ends with feature $i=8$ being acquired, which is the start of the complete trajectory). Trajectories belonging to the left-right transition in WS1 may be interpreted as joining the full right-left path. WS2 does not clearly disambiguate \Green{these dynamics -- it is not clear whether features 5-8 are acquired}. WS1, in the bottom right quadrant of the plot, shows \Green{some support for} the beginning of the complete progression beginning after the partial progression ends. \sifigten (iii) looks at two partial progressions again illustrating that in the grey region (acquisitions without support in the dataset), there can be a mixed signal from the two partial progressions.
\item \Green{{\em Noisy observations}: We consider the influence of noise in observations in \sifigeleven by looking at the single left-right progression conflated with noisy observations (from a cross-sectional dataset made up of 10 randomly sampled trajectories). From \sifigeleven (i)-(iii), the number of noisy (random acquisition of traits) observations increases, introducing breadth into the inferred posterior around the modal pathway (\sifigeleven for example). However, even with 50\% noisy observations in \sifigeleven (ii), it is possible to clearly recover the modal progression. \Sam{Even for the extreme case, the non-noisy pathway is almost exactly reproduced with the first greedy path across the hypercube.}}
\item {\em Repeated uniform sampling}: When repeated sampling occurs, it can strengthen the inference around where traits are acquired. For example, comparing the first four traits of \sifigten (i) and \sifigeleven (i), we can see that the repeated sampling afforded by 10 repeated trajectories almost completely removes any density for acquisition off the progression.
\item {\em Non-uniform sampling across the progression}: We consider this assumption in \sifigtwelve. When some states are sampled a greater number of times, parameterisations that lead to this state will have a stronger `signal' than those where the observation just occurs once. We illustrate this important effect with several examples. In all cases we consider the complete left-right progression but with the state $s = 11110000$ sampled 100 times more than the others. In \sifigtwelve (i) we see this state acts as a `gateway' by removing uncertainty for the acquisition of features present in $s$ after $s$ is encountered, and removing uncertainty in acquisition of features absent in $s$ before $s$ is encountered. In \sifigtwelve (ii), the right-left progression is also included but with uniform sampling. The non-uniform sampling leads to a much greater representation of the left-right progression. In \sifigtwelve (iii), two noisy trajectories are now included (only uniform sampling for the noisy trajectories). As the noise is uniform, acquisitions before $s$ still clearly resemble the progression, while features not present in $s$ become affected by the noise.
\end{itemize}

\section*{HyperTraPS and cancer progression models}\label{appendix:hypertraps_cpm}

Understanding pathways of cancer progression is highly complex due to widespread genetic heterogeneity at inter-patient, intra-patient and intra-tumour levels. Several methods aim to infer progression dynamics given different types and structures of observations \citep{Schwartz2017}. Additionally, cancer progression models can broadly be split into two classes: (i) approaches that consider the multitude of raw 'omic alterations that occur during carcinogenesis and (ii) approaches that take such alterations as absent or present (binarised data), and utilise description of the data at this level to consider progression. Our work fits within the second type of approach where relevant feature subsets have been identified and the presence of absence of such features is a measured aspect in samples.

For understanding variation between patients, no phylogenetic relationship is generally assumed to exist in the accumulation of genetic alterations. Key inference methods applied to binary data at the inter-patient level that determine feature relationships include Conjunctive Bayes network approaches \citep{Gerstung2009, Beerenwinkel2009, Gerstung2011, Montazeri2016} and the Tronco packages \citep{Loohuis2014,DeSano2016}, among a wide-range of similar approaches \citep{Beerenwinkel2015, Schwartz2017} and date back to oncogenetic tree models introduced by \citep{Desper1999}. Recent work by \citet{Diaz-Uriarte2018} suggests that, where complexity in the fitness landscape is present such as with the presence of reciprocal sign epistasis, Bayesian network type approaches in feature space may have shortcomings in being able to represent genetic pathways effectively due to the assumption of monotonicity. As we show in the main text, in contrast to other methods that work with absence/presence data, HyperTraPS focusses on the process of dynamic acquisition in the full space of binary states. This removes the restrictive prior assumption of monotonicity in feature relationships, while presenting tractable parameterisations  that include interactions between features. The HyperTraPS platform provides a new means for exploring oncogenetic data at large-scale, lifting this assumption.

In this article, we focus on inter-patient observations, for which established and well-studied datasets allow `benchmark' comparisons between approaches (as in the main text). However, we note that HyperTraPS' ability to infer dynamics from phylogenetically coupled observations also makes it an appropriate platform for the emerging field of intra-patient cancer study, where `phylogenetic' with somatic mutations as opposed to solely germline relationships between cells must be considered.
\Blueb{Recent methods for understanding feature relationships in single-cell data include SCITE \cite{Jahn2016} and SiFit \cite{Zafar2017}, while methods for relating the samples phylogenetically in single-cell data and evaluating clonal clusters include OncoNEM \citep{Ross2016}. \citet{Zafar2018} discuss these methods in the context of single cell cancer observations.
At the intermediate level of attempting to find common relationships in feature space across multiple cancer samples in different patients and different tissues, the recent Revolver platform attempts to provide a unifying interpretative approach via the method of transfer learning \cite{Caravagna2018}, and note that HyperTraPS could be readily applied to compilations of patient specific somatic trees too.} In Section \ref{section:tb} and Methos Appendix, we demonstrate that HyperTraPS allows efficient inference of many traits on phylogenies; application of HyperTraPS to these cancer `phylogenies', and comparison to these alternative approaches, will be the subject of future work.

\section*{Regularisation and model validation for ovarian and tuberculosis datasets}\label{section:regularisation_ovarian_tuberculosis}

\Sam{In STAR methods, we introduced a greedy backward selection process for inducing parsimonious parameterisations from samples of maximum likelihood models and demonstrated the process for an ensemble for the synthetic datasets (Fig. \ref{fig:cs12}D). In \sifigfourteen A, plots for the ovarian and tuberculosis datasets are also shown with the minimum AIC score at each $k$ from 1000 and 100 (for ovarian and tuberculosis respectively) unique greedy backward selection procedures for different maximum likelihood parameterisations. The AIC score is observed to decrease to a global minimum for each model. First-order models may only have a few parameters removed before reaching a minimum, while second order models, depending on the number of interactions in the underlying dataset, can have a greater proportion of parameters removed.}

\Blueb{For the ovarian dataset, the global minimum is sharply found at $k=30$ following an initial approximately linear decrease. Non-monotonic increase in AIC may then be seen, indicating the interacting nature of parameters to facilitate inference in this model, and is purely an artefact of the greedy backward selection process. For the $L=19$ genetic features of the complete tuberculosis dataset, a smoother increase in AIC is observed following a global minimum at $k=149$ parameters, indicating less strong direct interactions between parameter combinations.}

\Blueb{Validation calculations for the model (\sifigfourteen B(i)-(ii)) further support this message.
All models experience statistically significant support over the null model in terms of the log-likelihood ratio.
While the first order regularised model has improved predictive power over the null model, the second order regularised model provides around twice the increase in log-likelihood compared with the first order model. For the test dataset, the second order model has a marginal advantage over the first order model, both producing greater likelihoods than the null model. The lack of the same level of improvement from the second order model for the test dataset, indicates that the parameters remaining for the minimum AIC model from the validation set are not sufficient to capture the full heterogeneity of the datasets in these two specific cases.}

\section*{\pr{Analysis for specific biological datasets}}

\pr{For synthetic, CGH, and tuberculosis datasets, the original data naturally takes the form of presence/absence `barcodes' with defined features, and can therefore immediately be used in HyperTraPS.}

\pr{The TCGA study \citep{Bell2011} includes data on somatic copy-number alterations (SCNAs) from $N=489$ ovarian carcinoma DNA samples. The authors utilised a focal GISTIC methodology to identify significant peaks of amplification and deletion, and `key regions' of the genome where these SCNAs occurred. For a given observation, GISTIC analysis assigns an amplitude score and a significance level based on comparison to a control observation. We used these data to build a dataset describing whether or not a significant SCNA was found in each of $L=55$ chromosomal regions for each patient. We used the authors' GISTIC-derived magnitudes and significance levels, marking an SCNA as present in region $R$ if an observation was found overlapping with region $R$ for which the GISTIC magnitude exceeded $0.2$, the associated p-value was under a conservative genome-wide corrected value of $10^{-10}$, and the sign of the SCNA (deletion or amplification) agreed with that found in the original key region analysis. A range of changes in these thresholds for magnitude and significance did not have strong qualitative effects on the structure of the inferred pathways. For the PFG analysis with TCGA data we used the WS2 protocol as described in Methos Appendix.}

\pr{
In order to consider the data at different coarse grained levels from this full binary dataset, we created the following feature subsets:
\begin{itemize}
\item Chromosomal-level {\em TCGA-C1}: the union of presence/absence aberrations across a given chromosomal arm is considered, leading to $L=55$ chromosomal features. These are represented as chromosome number (integer), chromosome arm (p/q) and amplification or deletion (+/-)
\item High significance chromosomal-level {\em TCGA-C2}: where we consider the subset of chromosomal positions reported in Fig. 1c of \citet{Bell2011} in particular due to the authors indication that these were of greater significance. This led to a dataset with $L=27$ features.
\end{itemize}
}

\pr{We consider HyperTraPS and Bayesian network analysis of TCGA-C2 in the main text. In \sifigthirteen we demonstrate HyperTraPS inferences with TCGA-C1, across all chromosomal arms and key amplifications. Ordering histograms for the features for random walks with WS1 (blue) and WS2 (orange) are depicted with features ordered vertically by mean acquisition step. The inferred order of acquisition is highly heterogeneous, with early acquisitions observed in previously well known chromosomal regions (for example, {\em 8q+, 3q+, 5q-}). There is some multimodality observed in the WS1 and WS2 indicating multiple competing pathways. However, the dominant inferences are with respect to early and late acquisition at this large-scale level of description.
}

In \sifigfifteen , as described in the Main Text, we demonstrate the limited effect of phylogenetic structure in the tuberculosis dataset on the overall posterior structure.


\section*{Likelihood comparison of HyperTraPS with alternative Bayesian network approaches}\label{appendix:bayes-network-comparison}

In this section, we make a direct comparison of the likelihoods computed by the Bayesian network models compared with HyperTraPS. For the likelihoods to be comparable, we must include the additional probability of a random walk that leads to a target state emitting a signal in that target state by incorporating the $P_\text{emit}(\{0\}^L, t_i)$ for each $t_i \in D^\text{transitions}$. In this case, if signal emission is equally probable across all states, for every sample an additional factor of $1/(L+1)$ must be included for each sample given an irreversible walk from $\{0\}^L$ to $\{1\}^L$ may occupy $(L+1)$ states. This is discussed in further detail in \citet{Johnston2016}.

\sitableone provides a comparison of the maximum likelihood output of each model. HyperTraPS produces a similar maximum likelihood to the trained Bayesian network models for dataset $D_1$, while attaining greater likelihoods for datasets $D_2$ and $D_3$ from the ability to capture the competing pathways present in this dataset. For the ovarian dataset, the regularised (AIC criterion) maximum likelihoods are provided for Capri and HyperTraPS, while the maximum likelihood for CBN output is shown. HyperTraPS again attains the largest maximum likelihood. However, it should be noted that Capri model records a lower model complexity making the AIC scores of similar magnitude.

\section*{Additional interpretation of findings for tuberculosis dataset}\label{appendix:tb} \label{appendix:simmap}
\Sam{Additional comparisons can be made between the inferred order of polymorphism acquisition in Fig. \ref{fig:tb} and \sifignine B and the findings of by \citet{Casali2014}. Of the $L=19$ features used for the analysis, we pick a subset here that provide interesting discussion points with regard to co-associations discussed by the authors. These points demonstrate the ability of HyperTraPS to provide quantitative support for existing hypotheses, and to suggest new avenues of mechanistic research, in complex biological systems.
\begin{itemize}
\item {\em Drug-resistance and fitness compensatory mutations}: Of the $L=19$ features, the first 16 correspond to the drug-resistant polymorphisms within genes or in the promoter regions. The last three ({\em rpoA}, {\em rpoB} and {\em rpoC} are nonsynonymous SNPs within RNA polymerase genes. The authors considered the occurrence of compensatory mutations in {\em rpoA} and {\em rpoC} in response to drug-resistance polymorphism in {\em rpoB}. WS2 reveals an acquisition ordering with {\em rpoB} and {\em RRDR} being acquired prior to {\em rpoC}, suggesting a compensatory effect follows drug-resistance mutations in this case, while {\em rpoA} is acquired primarily in some cases and then typically later with similar acquisition patterns to {\em rpoC}.
  \item {\em Genetic sites particularly associated with adaptive selection}: Highly polymorphic genes conferring resistance are known to be {\em embB}, {\em pncA}, {\em ethA} \citep{Casali2014}. Interestingly these polymorphisms occur at a wide range of orderings within the inferred orderings, illustrative of their flexibility and why they may be particularly polymorphic -- they can play different roles in different progressions.
  \item {\em Transmissibility of drug-resistance}: With respect to transmissibility \citet{Casali2014} suggest that {\em katG} is prior to \emph{RRDR}, which is supported in the top two greedy paths highlighted in the hypercube plot in the main text Fig. \ref{fig:tb}.
\end{itemize}}

Here we make a direct comparison of the order in which mutations are acquired with Simmap, which takes the form of a continuous time Markov model with mater equation approach to acquiring characters that belong to leaves on a phylogeny. This approach runs into computational issues when the number of states under evolution grows large (only tractable in short run times for the tuberculosis up to $L\approx5$). This is in contrast to HyperTraPS which can handle the full $L=19$ traits.

As an illustration of compatibility with this alternative approach, we restrict the tuberculosis dataset to $L=3$ features ({\em katG}, {\em PinhA} and {\em RRDR}) with the full set of isolates and enforce single irreversible acquisitions as transitions within the Simmap model in order to make direct comparisons with HyperTraPS. In \sifigsixteen A, we show the output for the density of order of acquisition from simulated rate matrices outputted by Simmap with the hypercubic restriction imposed and irreversibility. Alongside in \sifigsixteen we show the result for WS2 with HyperTraPS (as the transitions performed with Simmap are to the sample data and do not fully acquire all features as is the case with WS1). The plots are \Green{in close agreement}, providing good validation that HyperTraPS generates results consistent with current platforms.

\newpage
\setcounter{page}{1}

\section*{HyperTraPS: Inferring probabilistic patterns of trait acquisition in evolutionary and disease progression pathways}
\section*{Supplementary Figures \& Tables}

\newpage

\begin{figure*}[!h]
  \centering
\includegraphics[width=\textwidth]{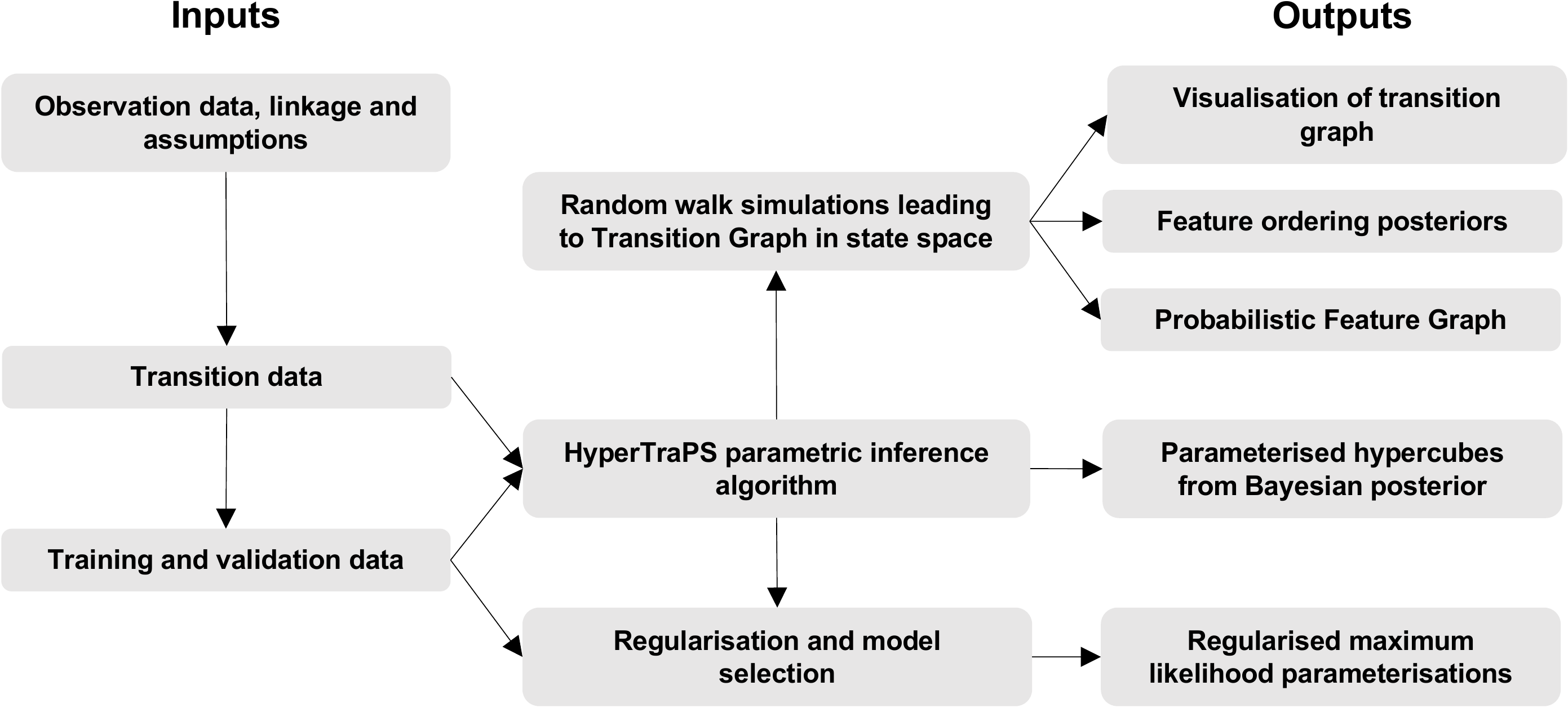}
\caption{(related to `HyperTraPS pipeline') {\bf An illustration of the pipeline from inputs to outputs with the underlying inference, application and description methods within HyperTraPS.}}
\label{fig:s1}
\end{figure*}

\begin{figure*}[!t]
\centering
\includegraphics[width=\textwidth]{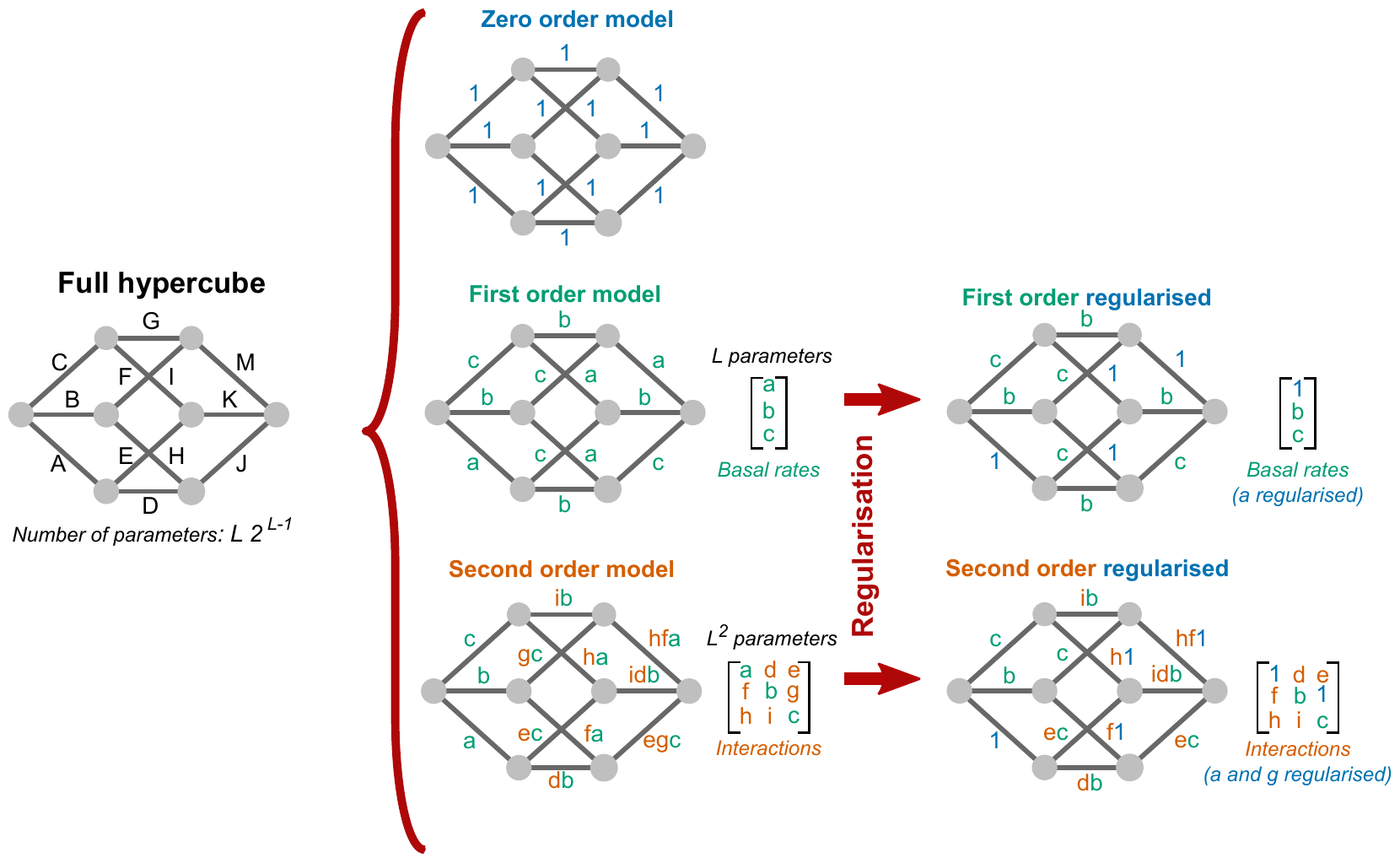}
\caption{ (related to `Tractable parameterisations of hypercube')
{\bf Tractable parameterisations and regularisation.}
A full irreversible directed hypercube is parameterised by edge set $W$ and contains $L2^{L-1}$ edges. We define three orders of model ({\em zero order}, {\em first order} and {\em second order}) for reducing the parameter space and regularised models ({\em first order regularised and second order regularised}). The zero-, first- and second- order models are nested in the sense that a second order model can capture the first order model (interaction terms all set to unity) and the first order model can capture the zero order model (all basal terms set to unity). In the example above, for $L=3$, the 12 edges of the full hypercube (A-K) are reduced down to combinations of a set of 9 parameters (a-i). The advantage becomes clear for larger $L$. At $L=16$, over 500,000 edge weights are reduced to just 256 parameters for the second order model. Regularisation harnesses structure in the data to further reduce model complexity. We utilise a greedy backward selection process to identify which parameters may be removed (set to the value of the zero order model, unity) and decrease a criterion, which we choose to be the Akaike Information Criterion. In the illustration above, for the first order regularised model, parameter $a$ is set to unity and, for the second order regularised model, parameters $a$ and $g$ are both set to unity  (as would be the case in a zero order model) with the consequent impact on the hypercube edge weights shown.}
  \label{fig:s2}
\end{figure*}

\begin{figure*}
  \centering
\includegraphics[width=\linewidth]{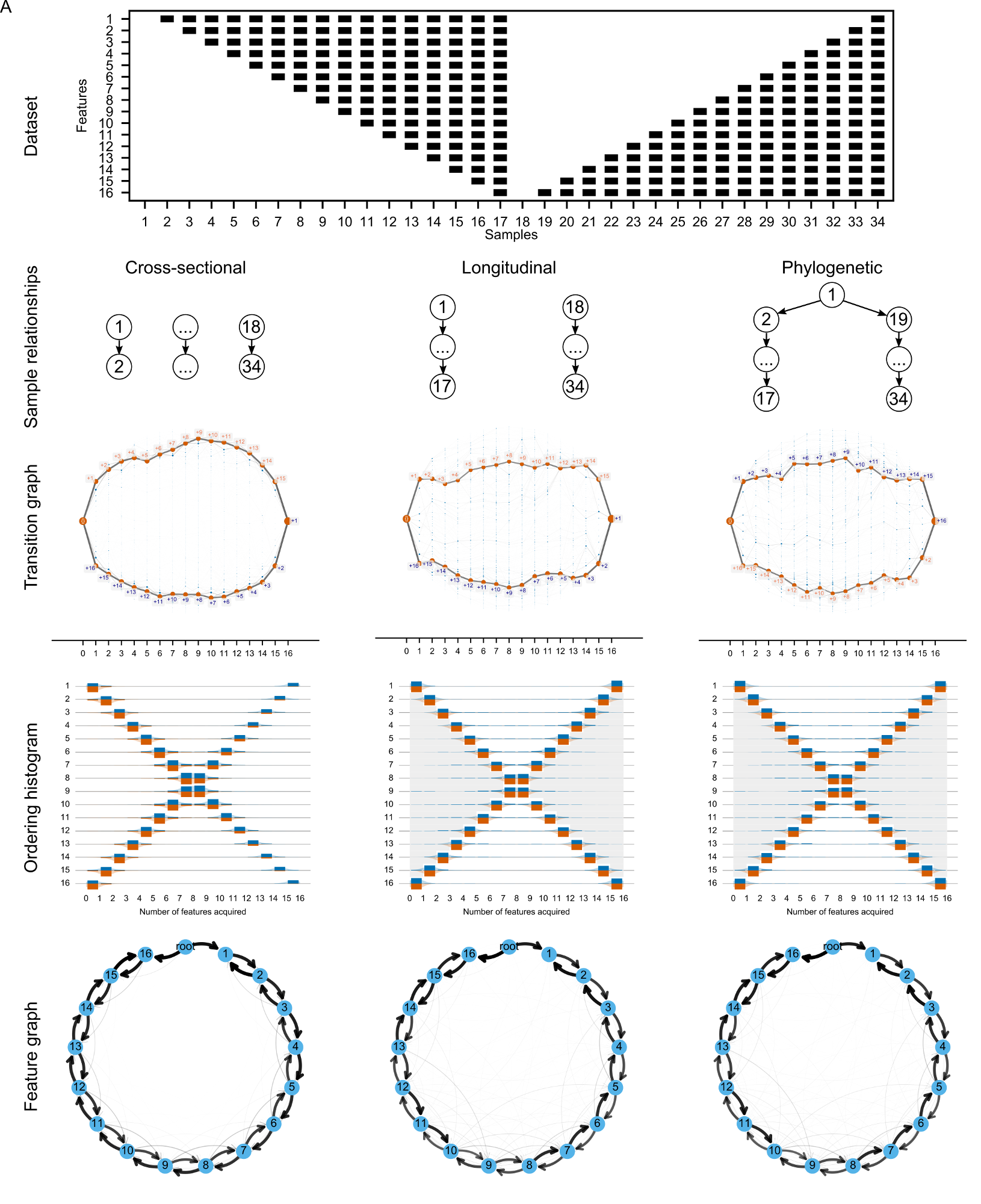}
\caption{(related to `Testing and validating HyperTraPS with differing data structures, volumes, and priors') \pr{\textbf{HyperTraPS inference with different data types.} The results of HyperTraPS inference using the $L = 16$ two-pathway synthetic system in the main text, where observations are cross-sectional, longitudinal, or phylogenetically linked. Pathways are readily recovered; posteriors are slightly sharper for cross-sectional data, as each observation is independent and thus provides more evidence than the coupled observations under the other two modes.}}
\label{fig:s3}
\end{figure*}

\begin{figure*}[!t]
  \centering
\includegraphics[width=\linewidth]{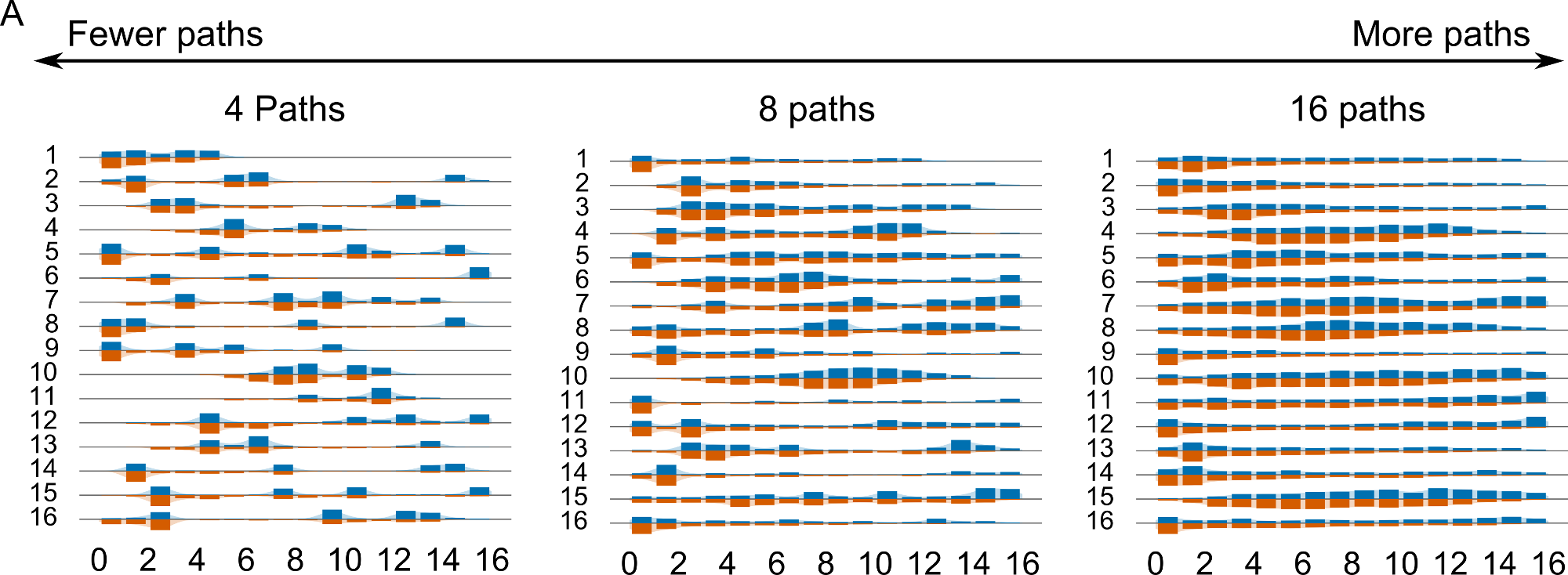}
\caption{(related to `Testing and validating HyperTraPS with differing data structures, volumes, and priors') \pr{\textbf{HyperTraPS inference with different numbers of competing pathways.} Posteriors corresponding to the hypercube plots in Fig. \ref{mtnewfig}A, using different $L = 16$ synthetic systems like those in the main text, but supporting different numbers $p>2$ of competing paths, with $N = 16p$ observations. Four and eight pathways are readily discerned; sixteen independent pathways poses more of a challenge, although posterior density is still aligned with the synthetic pathways. }}
\label{fig:s4}
\end{figure*}

\begin{figure*}[!t]
  \centering
\includegraphics[width=\linewidth]{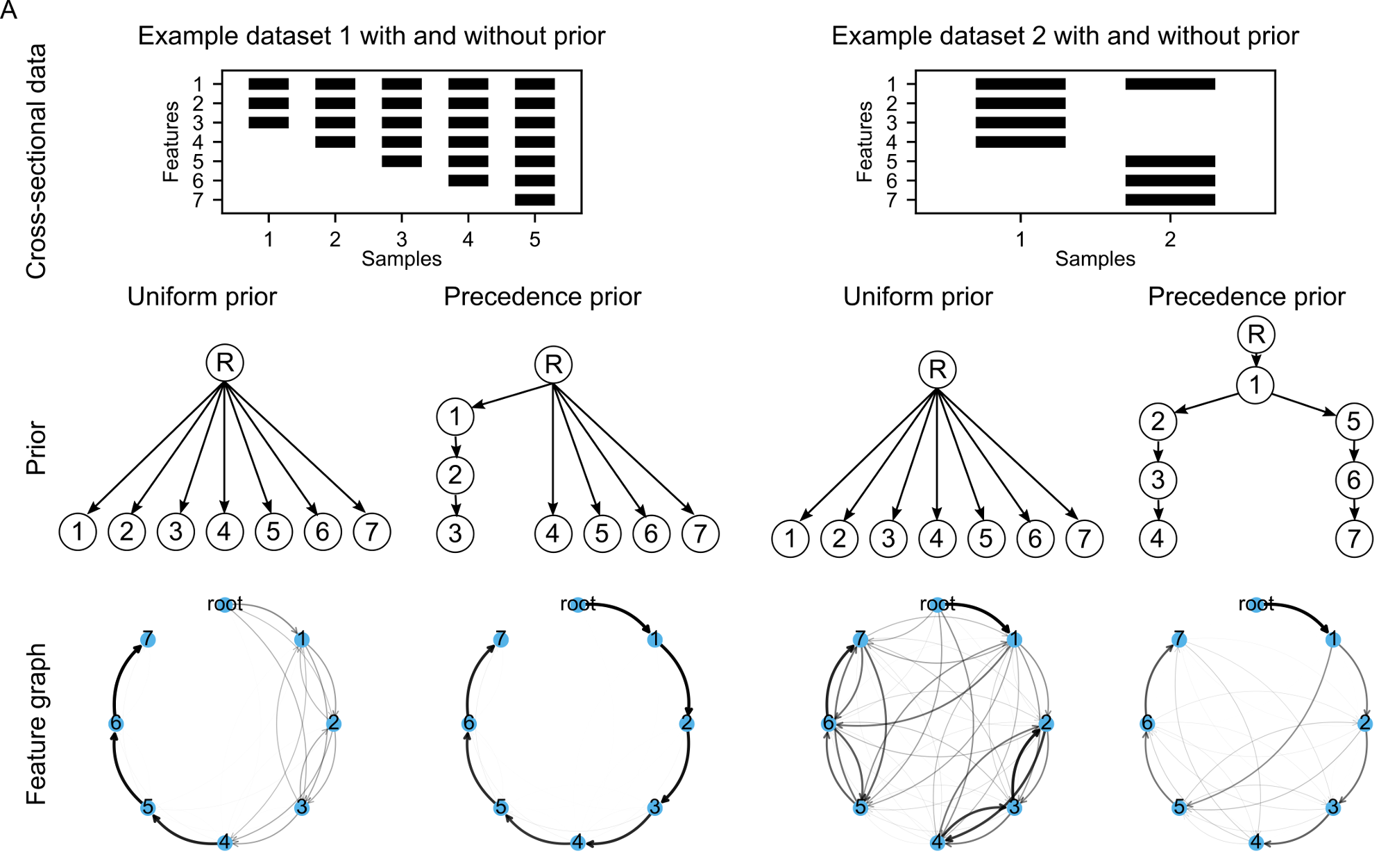}
\caption{(related to `Testing and validating HyperTraPS with differing data structures, volumes, and priors') \pr{\textbf{HyperTraPS inference including prior information on pathway structure.} Probabilistic feature graphs corresponding to the inclusion of prior knowledge in Fig. \ref{mtnewfig}C.}}
\label{fig:s5}
\end{figure*}

\begin{figure*}[!t]
  \centering
\includegraphics{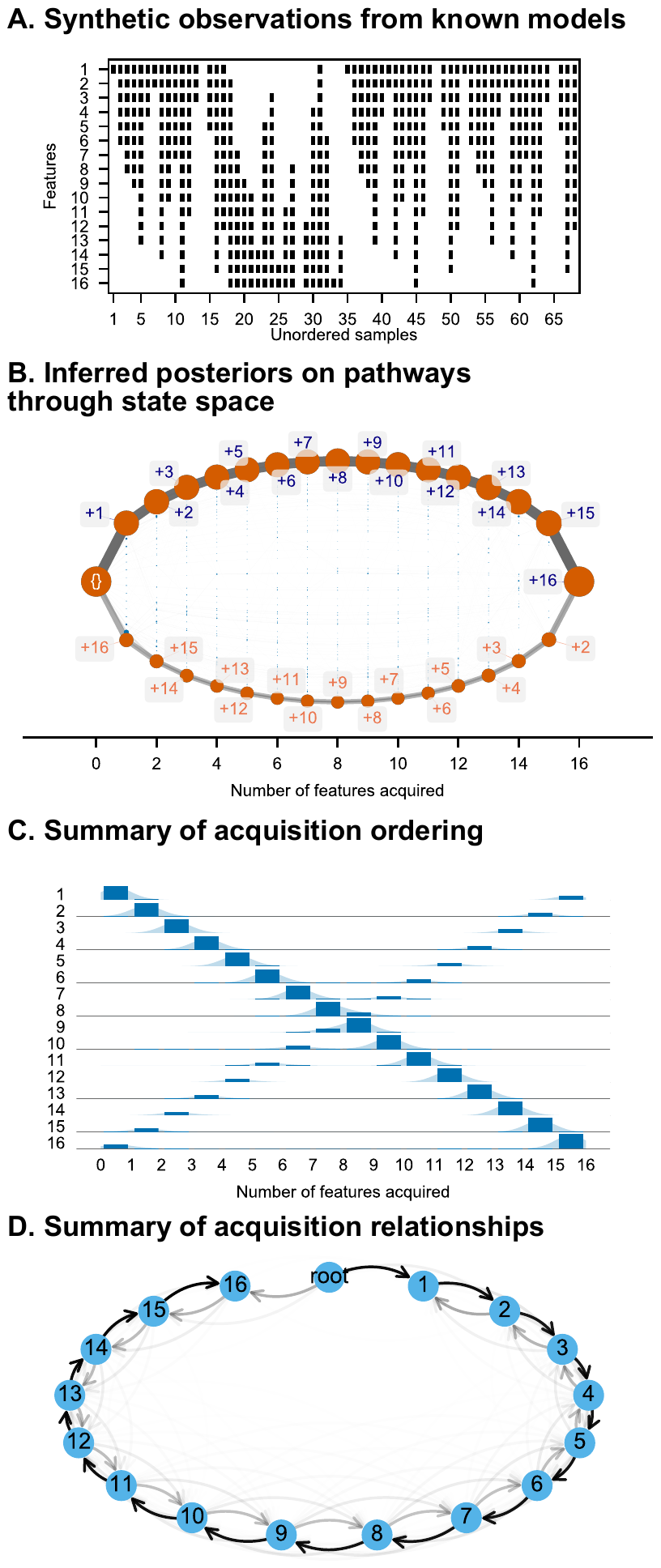}
  \caption{(related to `Additional synthetic cross-sectional dataset') {\bf HyperTraPS inference with additional synthetic dataset.}
  	\textbf{(A)} The structure of this synthetic dataset, supporting two competing pathways with features in different orders and with a likelihood ratio of 3:1 between the opposing orderings.
	\textbf{(B)} Inferred dynamics on the hypercubic transition graph. The two competing pathways are recovered in proportion to the amount they are observed in the dataset (the ratio of 3:1).
	\textbf{(C)} Inferred dynamics represented as the posterior probability that a feature (horizontal axis) is acquired at a given step (vertical axis), with bi-modality in proportion to the prevalence of each pathway.
	\textbf{(D)} Inferred dynamics represented as a graph summarising trait acquisition relationships. An edge from node $i$ to node $j$ suggests that trait $i$ is acquired in the previous step before the acquisition of  trait $j$. Again, two clear directions oaf acquisition can be seen with edge weights in proportion to their frequency in the underlying cross-sectional datasets.}
  \label{fig:s6}
\end{figure*}

\begin{figure*}[!t]
  \centering
\includegraphics[width=0.5\textwidth]{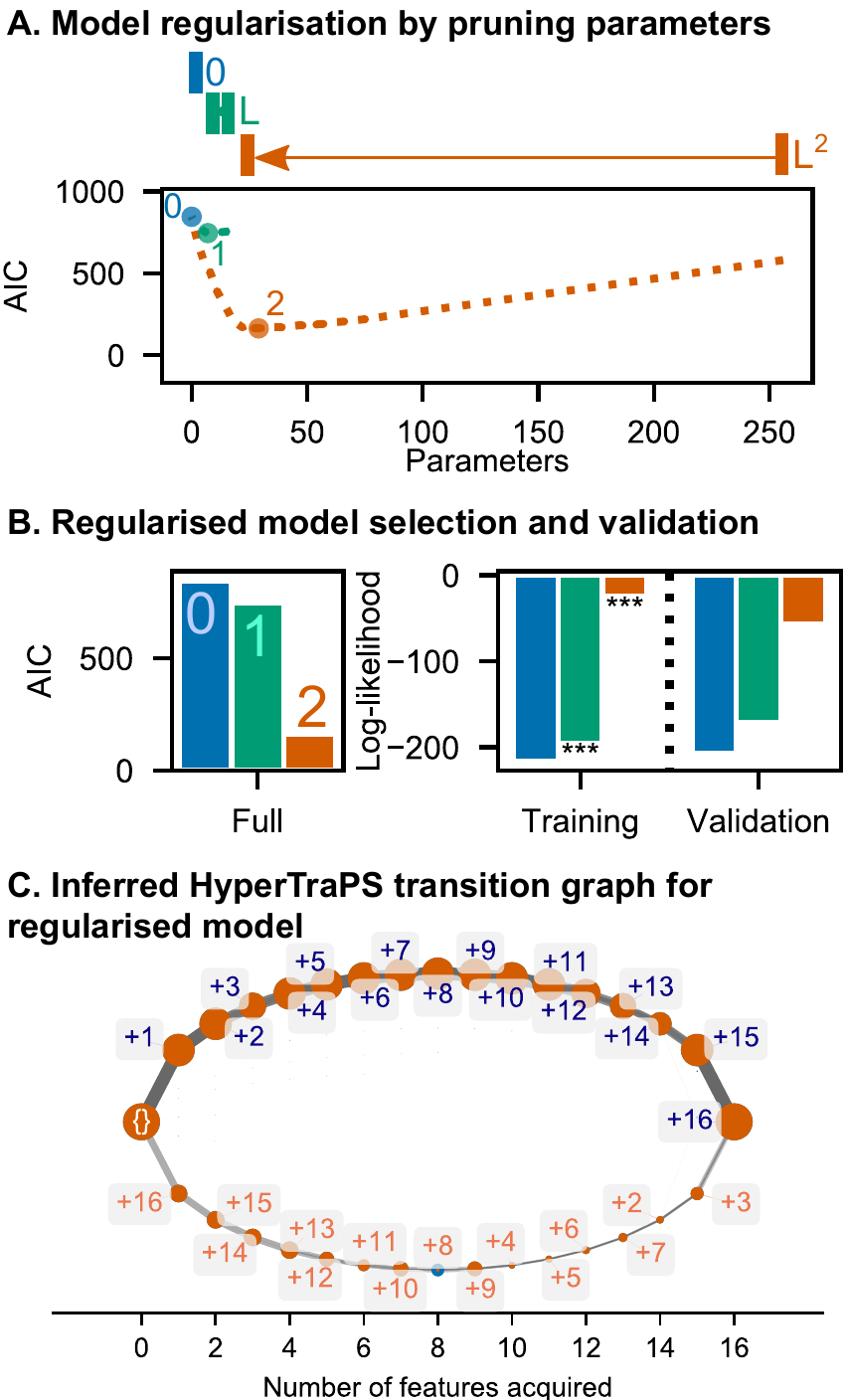}
  \caption{(related to `Additional synthetic cross-sectional dataset') {\bf Regularisation and validation for the additional synthetic dataset.}
  \textbf{(A)} Regularisation of the second order model (orange) leads to many fewer parameters than the full $L^2$ but still greater than the first order model's $L=16$ due to the two paths being present, necessitating interactions between features.
  \textbf{(B)} Regularised model selection and validation illustrates that the regularised first order model does better than the null model due to the full ordering of the dominant pathway that it is able to capture. The regularised second order model, however, results in much larger likelihoods still as it is able to capture both paths from the data.
  \textbf{(C)} Pathway structure remains well captured by the regularised model.}
  \label{fig:s7}
\end{figure*}

\begin{figure*}[!t]
    \includegraphics[width=0.48\textwidth]{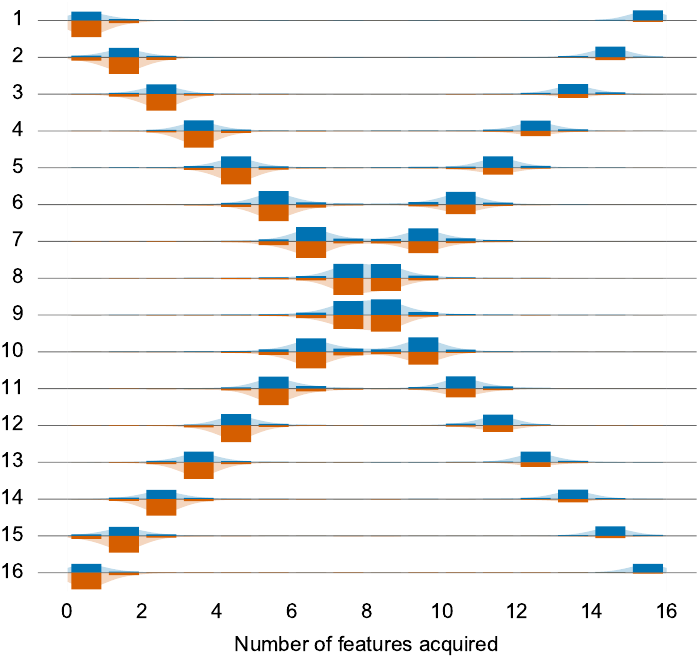}
    \includegraphics[width=0.48\textwidth]{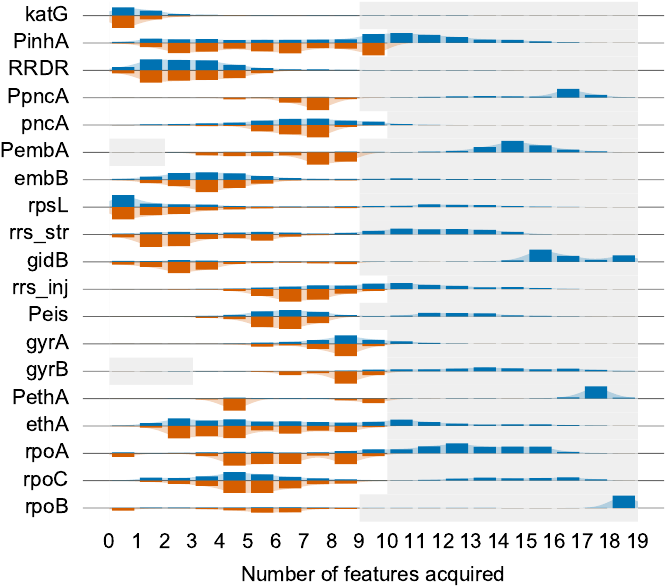}
    \caption{(related to `Alternative interpretation of inferred acquisition orderings') {\bf Comparison between WS1 and WS2 represented for the cross-sectional dataset (ii) (A) and tuberculosis (B) from the main text.} The blue bars are illustrate density corresponding to acquisitions with WS1 and the orange bars density for acquisitions with WS2. Kernel density estimates are overlaid to guide the eye.}
    \label{fig:s8}
\end{figure*}

\begin{figure*}[!t]
     \includegraphics[width=\textwidth]{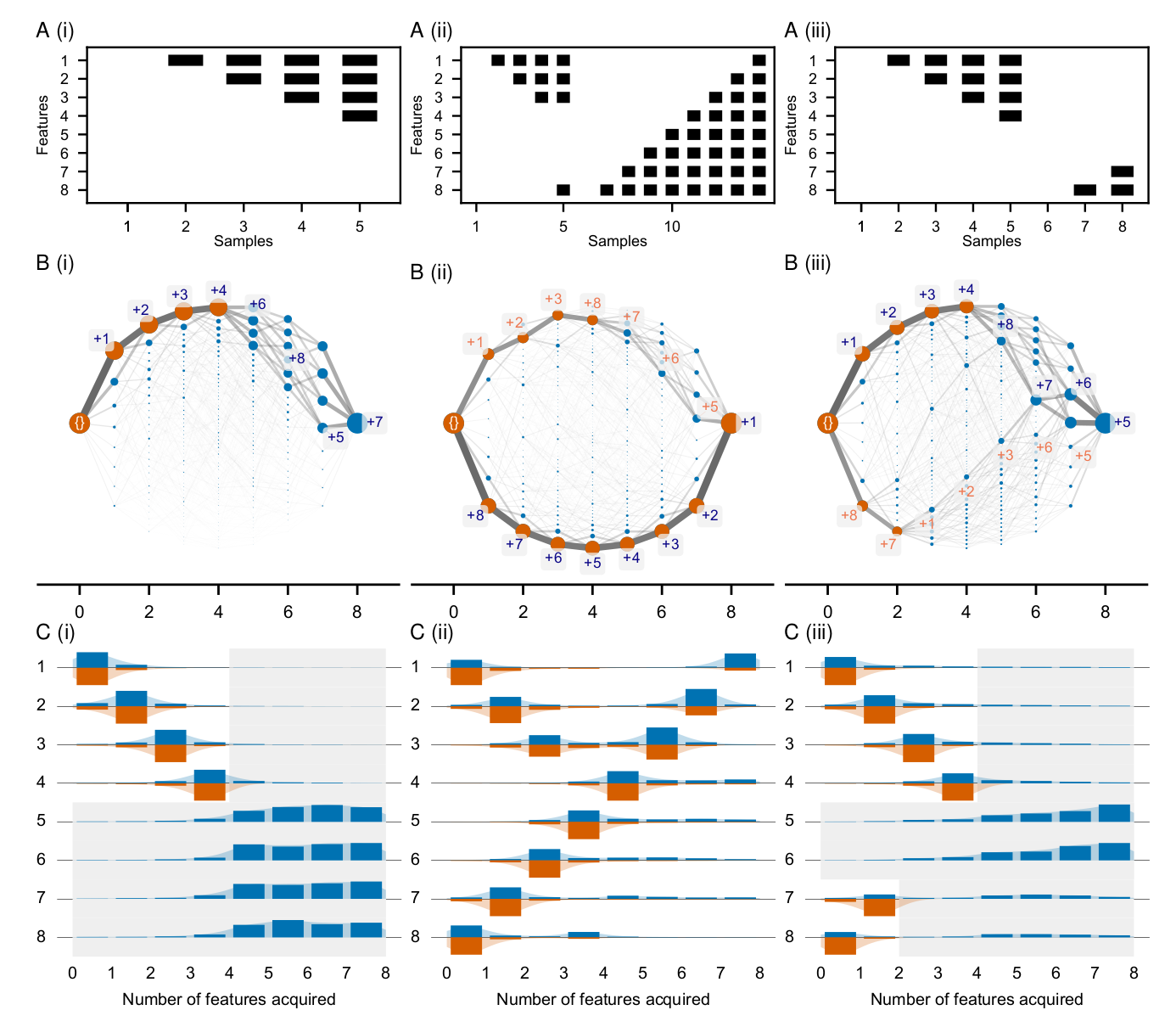}
 
  \caption{(related to `Implicit assumptions and interpretation of parameterisations') \Sam{{\bf HyperTraPS inference in the presence of partial and multiple progressions.} Three datasets are considered: (i) A single partial progression (1,2,3,4); (ii) A single partial progression (1,2,3,8) and a complete second progression (8,7,6,5,4,3,2,1); and (iii) Two partial progressions (1,2,3,4) and (8,7). In each case: \textbf{(A)} shows the dataset structure ({\em dataset plots}); \textbf{(B)} The inferred paths on the hypercube with samples from the second order posterior and WS1 simulations ({\em hypercube plots}). Orange vertices are observed in the dataset, while blue ones are not; and \textbf{(C)} The corresponding histograms for WS1 and WS2 ({\em histogram plots}).
For (i), the partial progression is inferred following by uniform acquisitions in line with the prior expectation. In the hypercube plots, paths on the hypercube are seen to diverge with equal proportion in this region illustrating this point.
For (ii), the hypercube plot highlights the ability to infer both progressions. The longer path has greater weight due to an increased number of observations associated. The greedily labelled paths show an interesting feature where at the end of the partial progression, as the last feature is the first feature of the complete progression, the pattern of acquisition seen in the second progression is `predicted' to occur in continued acquisition. This is visible in the histogram plot by the asymmetric density in WS1 flowing from feature $i=7$ for the fifth feature acquired onwards.
For (iii) with two partial progressions, the two paths are clearly distinguished in the hypercube plot with the same property of the progressions continuing on from each other after each partial progression is completed, eventually joining together after the sixth feature is acquired. The spread of other states encountered highlights the stochastic nature of the platform's predictions.}}
  \label{fig:s9}
\end{figure*}

\begin{figure*}[!t]
  \centering
       \includegraphics[width=\textwidth]{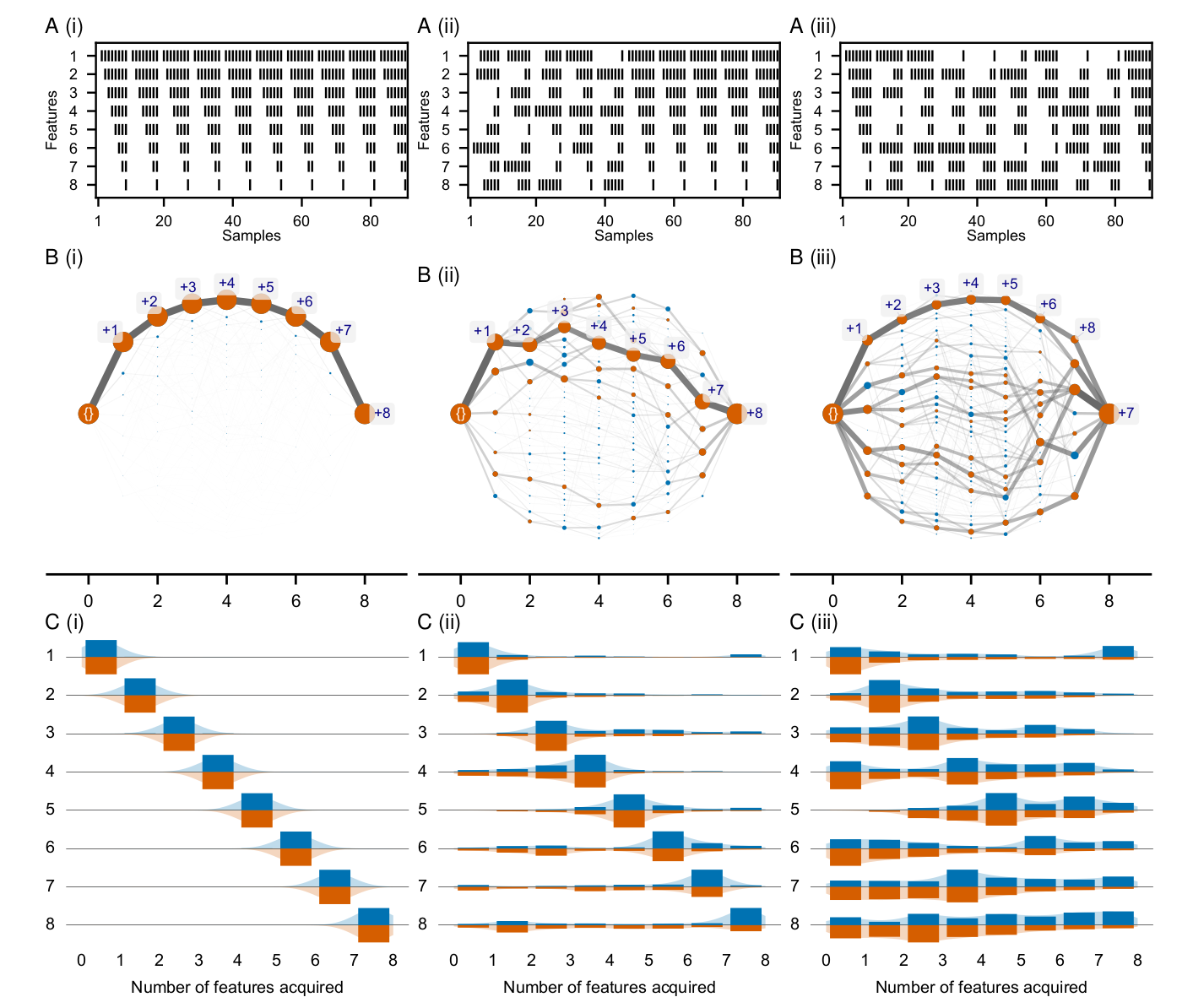}
    \caption{(related to `Implicit assumptions and interpretation of parameterisations') \Sam{{\bf HyperTraPS inference in the presence of noisy samples.} (i) One complete progression with ten samples from each state instead of a single sample ($|D|=10L$ compared to $|D|=L$). (ii) Five out of the ten trajectories part of the dataset involved the features being randomly acquired instead of the left-right progression. (iii) Nine out of the ten trajectories part of the dataset involved the features being randomly acquired instead of the left-right progression. The figure structure mirrors that of \sifigten.
  For (i), the hypercube plot and histogram plot shows more tightly defined paths due the ten-fold increase in data supporting the primary pathway, pushing the posterior towards the maximum likelihood parameterisation. In (ii), the introduction of this noise is visible but does not obscure the dominant non-noisy progression from being disambiguated. (iii) For (iii), the introduction of the uniform noise has a significant effect on the nature of paths observed across the hypercube, although even in this case it should be noted the appearance of the first greedy path being almost identical in structure to the non-noisy path structure.}}
  \label{fig:s10}
  \end{figure*}

\begin{figure*}[!t]
  \centering
       \includegraphics[width=\textwidth]{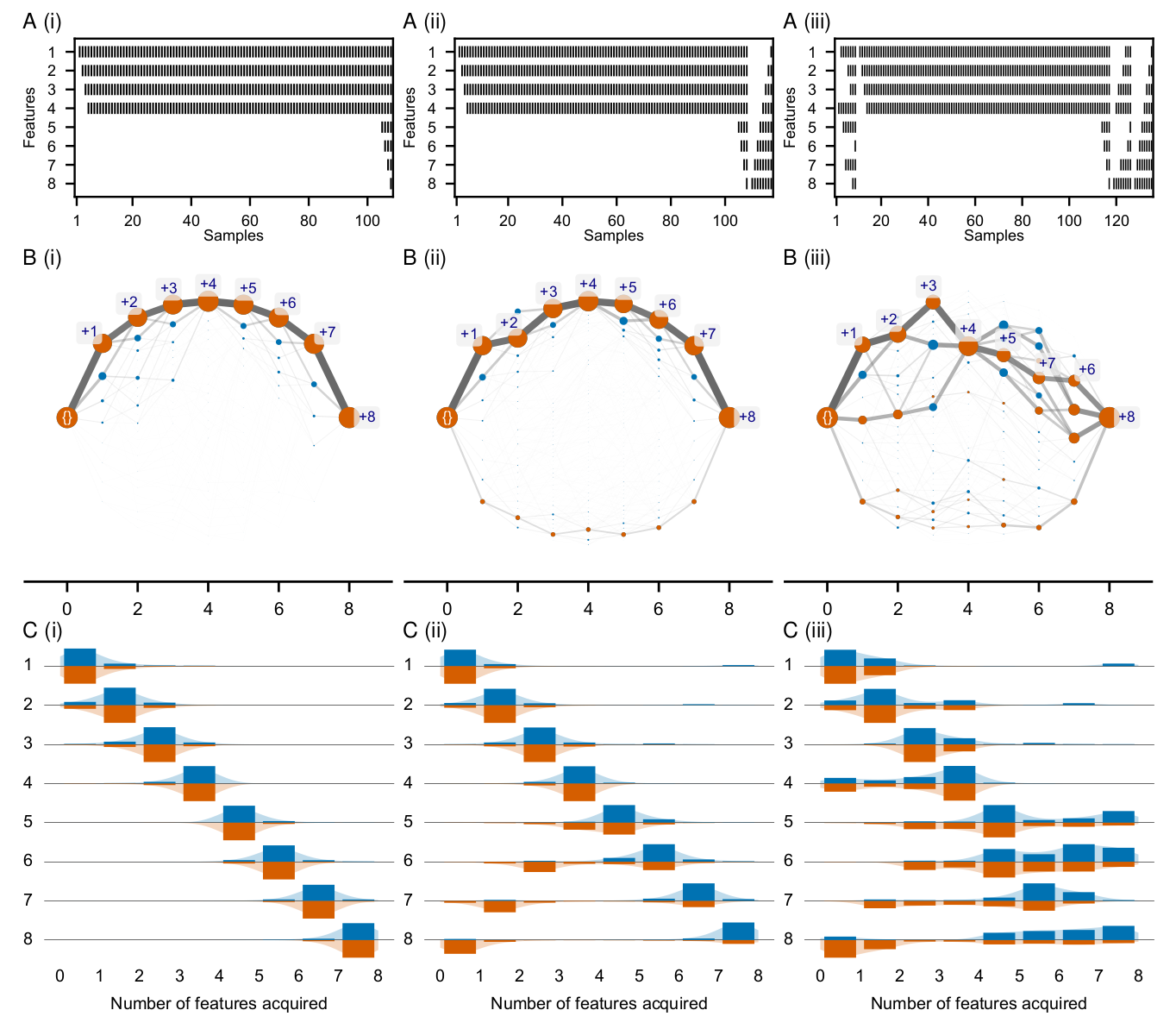}
    \caption{(related to `Implicit assumptions and interpretation of parameterisations') \Sam{{\bf HyperTraPS inference in the presence of non-uniform sampling.} In each of (i)-(iii) the state $s=11110000$ is sampled 100 times more than all other samples. (i) Only the single left-right progression. (ii) The single left-right progression with the non-uniform sampled middle state is present and a second progression with uniform sampling from right-left. (iii) Same as (ii) but a single noisy progression is added in each direction. The figure structure mirrors that of \sifigten and \sifigeleven.
For (i), the oversampled state acts as a gateway with uncertainty remaining in the regions where acquisition occurs before and after the gate. For example, $f_{45}\approx 0$ in contrast to \sifigten (a), while $f_{43}\neq 0$ as for the uniform case.
For (ii), where two progressions are present but only the left-right has oversampling in the middle, due to the oversampling in the left-right path there is a large bias towards random walks from $0^L$ following this path, as seen by the strength of corresponding path in the hypercube plot. WS2 allows for this to be accounted for illustrating the other pathway more clearly as the simulations ensure the right-left progression is visited.
For (iii), noise is now introduced for both progressions.
As the noise is uniform, acquisitions before the oversampled state $s$ still resemble the dominant progression, while subsequently the noise clearly affects the order of acquisition increasing the uniformity of feature acquisition. The right-left progression becomes difficult to distinguish at all due to a lack of random walks beginning at $0^L$ following this progression. However, the ability for the inference to perform random walks that take this weaker and noisy second progression is remarkable as observed by the fact orange states from the data associated with the progression are still encountered.}}

  \label{fig:s11}
\end{figure*}

\begin{figure*}[!h]
  \centering
    \includegraphics[width=\textwidth]{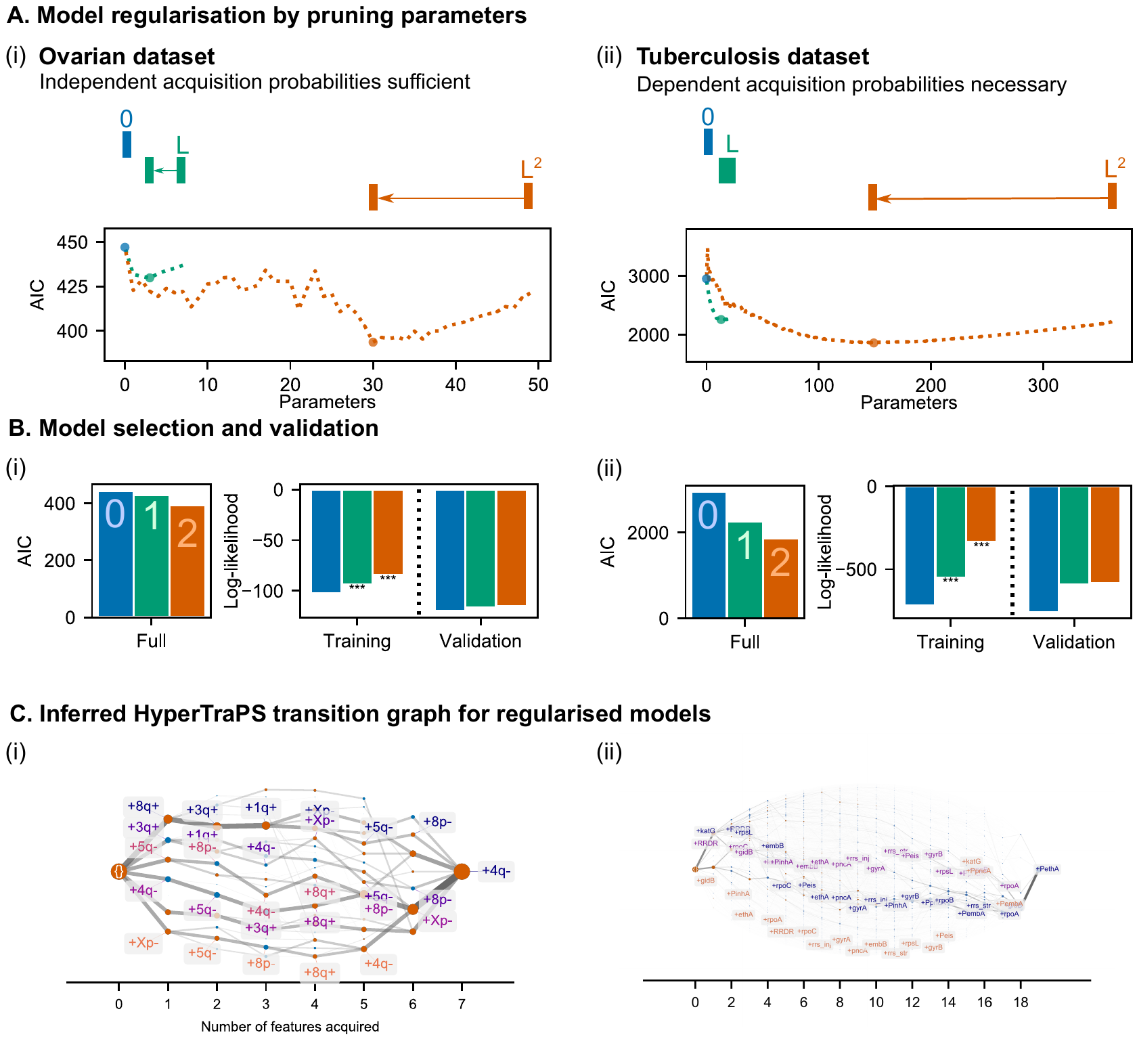}
  \caption{(related to `Regularisation and model validation for ovarian and tuberculosis datasets') {\bf Regularisation and model validation for the ovarian and genetic tuberculosis dataset.} \textbf{(A)} Regularisation of the parameterisations for the (i) ovarian dataset and (ii) tuberculosis genetic dataset ($L=19$ genetic sites). Dashed green and orange lines illustrate the minimum AIC found at each value of $k$ over the ensemble of backward selection processes. Circles illustrate the minimum for each order of model. The second order model is favoured produces lower AIC scores in both cases. \textbf{(B)} Model validation for the (i) ovarian dataset and (ii) tuberculosis dataset. In each of B(i) and B(ii), the left-hand plot depicts lower AIC scores for the second order models. The right-hand plots show highly significant second order regularised models compared to the null model and much larger log-likelihoods on the validation datasets. \Blueb{\textbf{(C)} Transition graphs constructed from WS1 random walks with the minimum AIC second order regularised models for each dataset.}} 
  \label{fig:s13}
\end{figure*}

\begin{figure}[!t]
  \centering
\includegraphics[width=\linewidth]{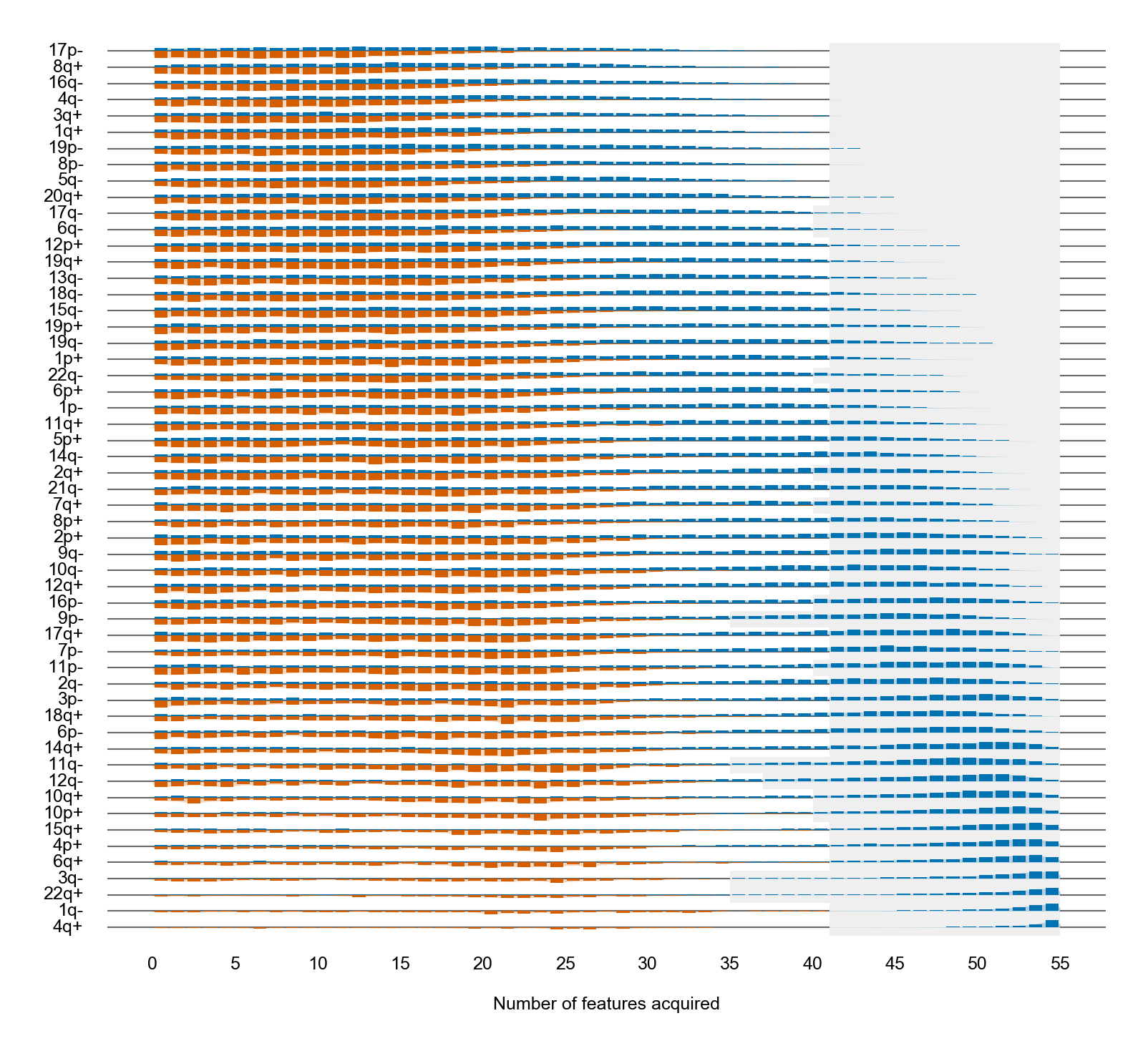}
\caption{(related to `Analysis for specific biological datasets') \pr{\textbf{Ordering histograms for random walks from posterior samples for the TCGA-C1 dataset are depicted.} Random walks with WS1 (blue) and WS2 (orange) are summarised into feature acquisition proportions at a given time. The features are ordered by mean acquisition time from WS1. The order of acquisition is highly heterogeneous, with general trends of early and late acquisition being clearly attributable to each feature. However, there is wide dispersion in the exact time of acquisition in almost all cases. There is some multimodality observed in the WS1 and WS2 indicating multiple competing pathways.}}
\label{fig:s12}
\end{figure}

\begin{figure*}[!t]
  \centering
\includegraphics[width=\linewidth]{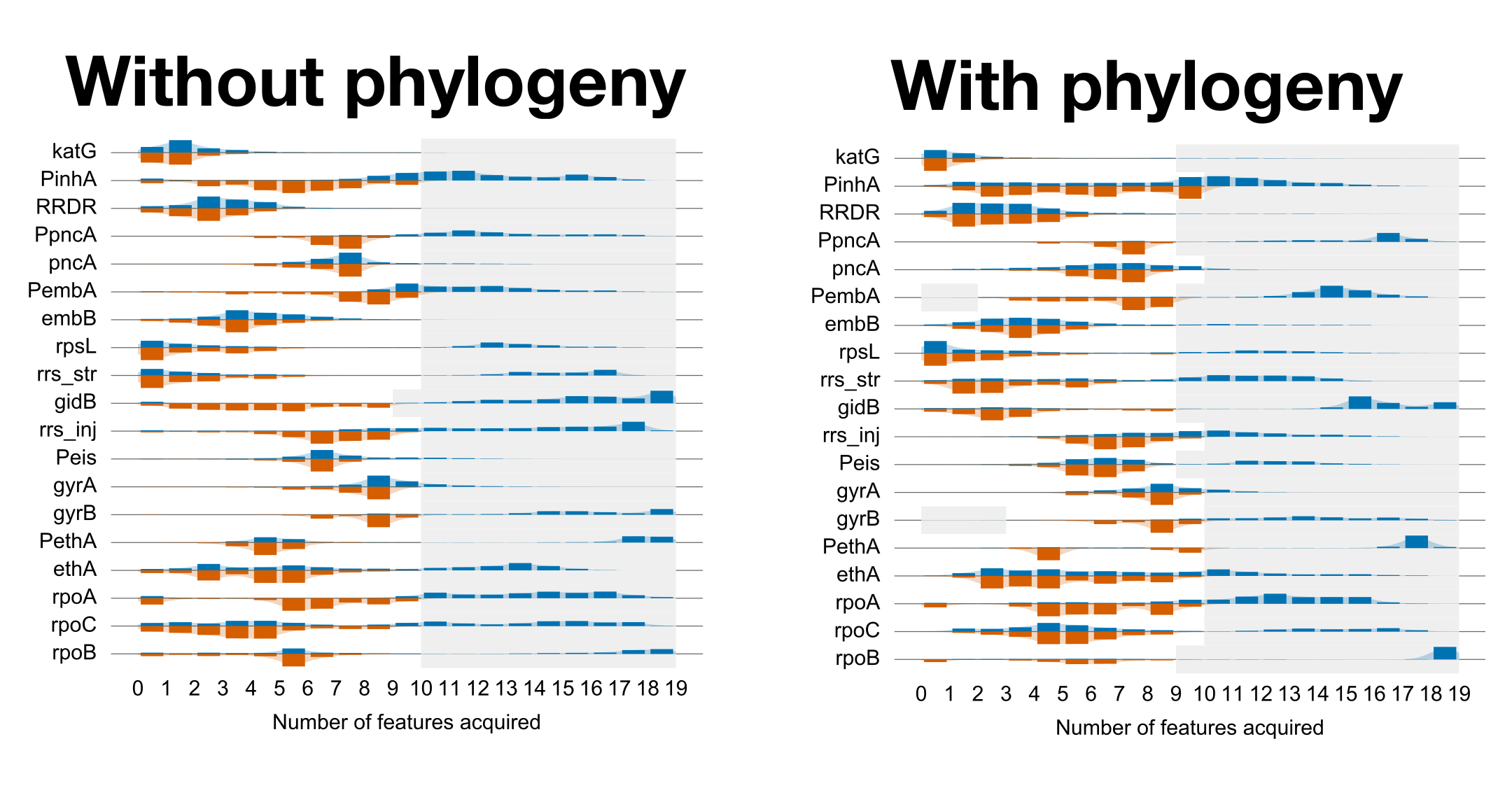}
\caption{(related to `Additional interpretation of findings for tuberculosis dataset') \pr{\textbf{Tuberculosis pathway inference and phylogenetic information.} (left) The inferred structure of tuberculosis feature acquisitions, given the phylogeny used in the main text. (right) The inferred structure in the absence of the phylogeny, treating each observation as independent. Most ordering posteriors remain qualitatively similar to those inferred with phylogenetic information, illustrating their robustness to errors in phylogenetic structure.}}
\label{fig:s14}
\end{figure*}

\begin{figure*}
 \centering
     \includegraphics[width=0.45\textwidth]{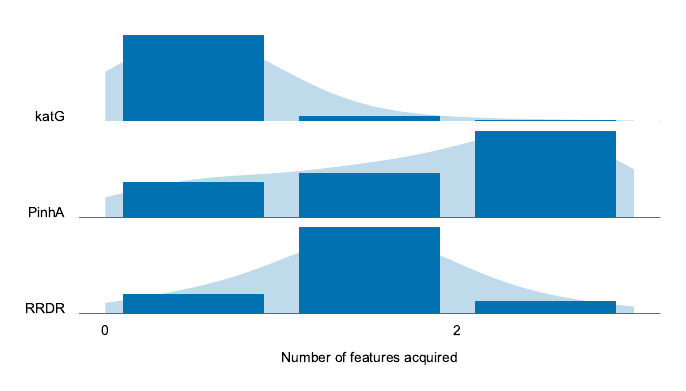}
     \includegraphics[width=0.45\textwidth]{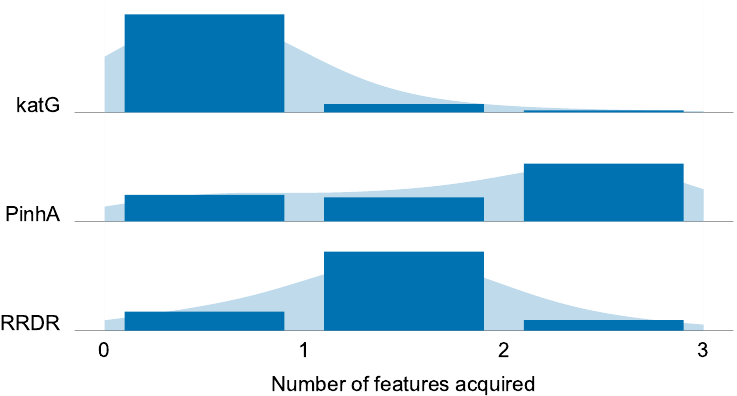}
 
   \caption{(related to `Additional interpretation of findings for tuberculosis dataset') Comparison of the tuberculosis dataset analysed with both HyperTraPS (A) and Simmap (B) on the restricted, tractable set of genetic sites: {\em katG}, {\em PinhA} and {\em RRDR}.}
   \label{fig:s15}
\end{figure*}

\begin{table}[!t]
  \centering
  \begin{tabular}{ p{3cm} p{4cm} p{4cm} p{4cm} }
    \hline
    Dataset & Maximum regularized likelihood with Capri & Maximum likelihood with CBN & Maximum likelihood with HyperTraPS \\
    \hline
    Synthetic $D_1$  &  -16.64  &   -17.52 & -16.64   \\
    Synthetic $D_2$  &  -46.97  &  -48.96  & -41.59   \\
    Synthetic $D_3$  &  -88.81  &  -86.05  & -80.04   \\
    Ovarian CGH & -356.57* & -380.01 & -347.72* \\
    \hline
  \end{tabular}
  \caption{{\bf Maximum likelihood values for Capri, CBN and HyperTraPS outputs with each synthetic cross-sectional dataset and ovarian CGH dataset.} Where there is a single progression (dataset $D_1$) all models reproduce the similar maximum likelihoods. Where there is more than a single progression (datasets $D_2$ and $D_3$), the additional stochastic flexibility available in HyperTraPS parameterisations allows models with larger maximum likelihoods to be recovered. For the ovarian dataset, HyperTraPS and Capri both have likelihoods compared in regularised forms (denoted with asterisks), with HyperTraPS again attaining the largest likelihood. It should be noted however, that the model complexity of the Capri model is less than that for HyperTraPS in this case, leading to a lower AIC score (not shown above).
  }
\label{table:s1}
\end{table}

\end{document}